\documentclass{article}
\usepackage{graphicx}  % standard LaTeX graphics tool;
\usepackage{amsmath}   % For getting proper math eqns
\usepackage{amssymb}   % Used for getting various symbols eg. gtrsim, lesssim etc.
\usepackage{bm} % For bold math (esp of lower greek letters)
\usepackage{dcolumn}% Align table columns on decimal point
\usepackage{color}
\usepackage{mathrsfs}
\usepackage{amsfonts}
\usepackage{varioref}
\usepackage{subcaption}
\usepackage{slashed}
\RequirePackage[colorlinks,citecolor=blue,urlcolor=magenta,linkcolor=blue]{hyperref}
\usepackage{comment}
\def\be{\begin{equation}}
\def\ee{\end{equation}}
\DeclareMathOperator{\sech}{sech}

\labelformat{section}{Section #1} 
\labelformat{subsection}{Section #1} 
\labelformat{subsubsection}{Section #1}
\labelformat{subsubsubsection}{Section #1}
\labelformat{equation}{Eq.~(#1)} 
\labelformat{figure}{Fig.~#1} 
\labelformat{subfigure}{Fig.~\thefigure#1} 
\labelformat{table}{Tab.~#1} 
\labelformat{appendix}{Appendix #1}
\title{\bf Fermionic entanglement in the presence of background electric and magnetic fields}
\author{Shagun Kaushal\footnote{shagun123@iitd.ac.in}\\
\small{Department of Physics, Indian Institute of Technology Delhi, Haus Khas, New Delhi 110 016, India}\\}
\date{ }  %% This command will suppress printing the date. 
\begin{document}
\maketitle
\begin{abstract}
\noindent
In this study, we investigate the fermionic Schwinger effect in the presence of a constant magnetic field within $(1+3)-$dimensional Minkowski spacetime, considering both constant and pulsed electric fields. We analyze the correlations between Schwinger pairs for the vacuum and maximally entangled states of two fermionic fields. The correlations are quantified using entanglement entropy and Bell's inequality violation for the vacuum state, while Bell's inequality violation and mutual information are used for the maximally entangled state. One can observe the variation of the entanglement produced for fermionic modes with respect to different parameters. Additionally, we discuss the key differences from the behaviour of scalar fields in this context. This study offers deeper insights into quantum field theory and the dynamics of entanglement in the fermionic Schwinger effect.
 
 \end{abstract}
\newpage
\tableofcontents
%%%%%%%%%%%
\section{Introduction}\label{S1}
Quantum entanglement is even more mysterious than standard quantum mechanical processes. Experimental observations, such as the violation of Bell inequalities, which cannot be explained by classical theories relying on local hidden variables, have firmly established quantum entanglement on strong physical grounds ~\cite{bell_1, bell_2, bell_3, bell_4, Aspect1, Aspect2}.
Moreover, entanglement has numerous promising applications across diverse fields, including quantum communication and teleportation, quantum cryptography, and quantum computing. It also plays a crucial role in black hole thermodynamics \cite{Mukohyama:1996yi, MartinMartinez:2010ar} and the information loss problem \cite{Horowitz:2003he, Alsing:2006cj, Adesso:2007gm}, prompting extensive research on the generation and degradation of entanglement across various systems.

% These investigations cover entanglement in both inertial \cite{Peres:2002ip, HSSS} and non-inertial frames \cite{Alsing:2003es, FuentesSchuller:2004xp}, as well as its creation in expanding spacetime \cite{Fuentes:2010dt, Kanno:2014bma, Vennin, SSSS}. 

A natural framework for studying quantum entanglement involves systems where pair creation occurs. In quantum field theory, the vacuum is characterized by fluctuations that manifest as virtual particle-antiparticle pairs \cite{Reynaud:2001kc, Sakharov:1967pk, Streeruwitz:1975wzf, Zeldovich:1971mw, Mainland:2018yqz}. These pairs can be separated by the energy of external fields or the causality of spacetime \cite{Mainland:2018yqz, Kim:2016xvg, Srinivasan:1998fk}. For example, when virtual pairs are electrically charged, such as positrons and electrons, a sufficiently strong background or classical electric field can tear these virtual pairs apart, allowing them to be observed as real particle-antiparticle pairs—a phenomenon known as the Schwinger effect \cite{Avramidi:1989ik, Schwinger1, Sauter, Schwinger}. Pair creation can also occur due to the presence of a time-dependent gravitational field \cite{Parker:1968mv, Parker:1969au} or in certain non-trivial backgrounds \cite{QFTCS}. Hawking radiation from a collapsing black hole is another example of pair production and vacuum instability in a non-trivial background \cite{Hawking, Hawking1}.

The entanglement properties between particle-antiparticle pairs created by a background electric field have been investigated in flat spacetime for both scalar and fermionic fields, as shown in \cite{Li:2016zyv, Wu:2020dlg, Ebadi:2014ufa, Adorno:2015ibo}. These studies revealed that entanglement increases with the strength of the electric field for the vacuum state. Entanglement generated by time-dependent gravitational fields has also been explored, as seen in \cite{Fuentes:2010dt, Arias:2019pzy, EE, bell:2017, QC in deSitter}. The entanglement properties between particles created due to background electric and gravitational fields were previously studied in \cite{Ebadi2015} and for holographic aspects of entanglement, we refer our reader to \cite{Grieninger:2023pyb, Grieninger:2023ehb}.

These instances involve pair creation due to background electric or time-varying gravitational fields. In the context of the Schwinger effect, it is natural to consider both electric and magnetic fields, especially in scenarios such as primordial fields in the early universe or fields surrounding astrophysical black holes. The influence of a magnetic field on particle-antiparticle entanglement in the presence of an electric field is particularly intriguing. Studies on this effect have been conducted in Minkowski spacetime, within inflationary scenarios, and from the perspective of a Rindler observer, as discussed in \cite{HSSS, SSSS, SK}.

In \cite{HSSS}, the impact of a background magnetic field on the correlations between particles and antiparticles produced by a constant background electric field was analyzed for a complex scalar field in Minkowski spacetime. In this paper, we shift our focus to conducting a similar analysis for more realistic fermions. Additionally, we explore the scenario involving time-dependent electric fields.

The rest of this paper is organized as follows: In \ref{Fermionic field in background electric and magnetic field}, we quantize the Dirac field in the presence of background electric and magnetic fields and derive the Bogoliubov relations between the modes. The \ref{Correlation measures} provides a brief review of the relevant correlation measures. In \ref{ND}, we compute the number density and correlations for the vacuum state. Next, in \ref{Maximally entangled state}, we analyze the correlations in various sectors of a maximally entangled state. In \ref{Fermions coupled to time-dependent electric field and constant magnetic field}, we extend this analysis to include a time-dependent electric field. Finally, we present our summary and outlook in \ref{Summary and outlook}.

We shall work with the mostly positive signature of the metric in $(3+1)$-dimensions and will set $c=k_B=\hbar=1$ throughout.
\section{Dirac fermions coupled to background electric and magnetic fields of constant strengths}
\label{Fermionic field in background electric and magnetic field}
We consider a fermionic field in the presence of background electric and magnetic fields of constant strength. Our analysis here is parallel with \cite{ Ebadi:2014ufa, SSSS}. The Dirac equation coupled with a gauge field in the $(1+3)$-dimensional Minkowski spacetime is given by
\begin{equation}
\label{DiracMinkowski}
    \left(i \gamma^{\mu} \partial_{\mu}-q \gamma^{\mu} A_{\mu}-m\right) \Psi(x)=0
\end{equation}
where $\gamma^{\mu}$ and $\Psi(x)$ are the Dirac matrices and spinors, respectively. We consider the external gauge field the same as the scalar field case, i.e., $A_\mu = (-Ez,-By,0,0)$ \cite{HSSS}.
We introduce $\chi(x)$ as
\begin{equation}
\label{G}
    \Psi(x)=\left(i \gamma^{\nu} \partial_{\nu}-e \gamma^{\nu} A_{\nu}+m\right) \chi(x)
\end{equation}
On substituting \ref{G} into \ref{DiracMinkowski}, the second-order differential equation we obtain is
\begin{equation}
\label{2ndorder}
    %\left(i \gamma^{\mu} \partial_{\mu}-e \gamma^{\mu} A_{\mu}-m\right)\left(i \gamma^{\nu} \partial_{\nu}-e \gamma^{\nu} A_{\nu}+m\right)G(x)=
\left[\left(\partial_\mu+ieA_\mu\right)^2-m^2-i\frac{e}{2}\gamma^\mu \gamma^\nu F_{\mu \nu}\right]\chi(x)=0
\end{equation}
where $F_{\mu \nu}$ is the field strength tensor. Using the expression for $A_\mu$ and the ansatz $\chi(x)= e^{-i(k^0t-k^x x)}\chi^{(p)}_{s}(y,z)\epsilon_s$, for the particle mode ($p$ stands for particle), the above equation becomes
\begin{equation}
	\Big[
	-(
	k^0 + eEz
	)^2
	-
	(
	k^x - eBy
	)^2
	+
	\partial_y^2
	+
	\partial_z^2
	-
	m^2
	-ie(\gamma^0\gamma^3E+\gamma^1\gamma^2B) \Big]
	\chi^{(p)}_{s} (y,z)\epsilon_s
=
	0
\label{eq:EOM1}
\end{equation}
Since the matrices $\gamma^0\gamma^3$ and $\gamma^1\gamma^2$ commute, we may treat $\epsilon_s$ to be their simultaneous eigenvectors given as
\be
\begin{aligned}
\label{epsilon_s}
&  \epsilon_{1}=\left(\begin{array}{l}
1 \\
0 \\
1 \\
0
\end{array}\right), \quad \epsilon_{2}=\left(\begin{array}{c}
0 \\
1 \\
0 \\
1
\end{array}\right),
& \epsilon_{3}=\left(\begin{array}{c}
0 \\
1 \\
0 \\
-1
\end{array}\right), \quad
 \epsilon_{4}=\left(\begin{array}{c}
1 \\
0 \\
-1 \\
0
\end{array}\right).
\end{aligned}
\ee
These eigenvectors correspond to the eigenvalues $\lambda_s=\pm 1$ and $\beta_s=\pm i$ corresponding to $\gamma^0\gamma^3$ and $\gamma^1\gamma^2$, respectively. For spin-up and spin-down particle modes, $s=1$ and $s=2$, respectively, and for spin-up and spin-down antiparticle modes, $s=3$ and $s=4$, respectively. Next, we decouple \ref{eq:EOM1} using $\chi^{(p)}_{s}(y,z)=h_s(y) f^{(p)}_s(z)$ and decoupled equations read as
\begin{eqnarray}
\left(\partial_z^2-(k^0+eEz)^2-m^2-ieE\lambda_s-S_s\right)f^{(p)}_s(z)=0\\ \left(\partial_y^2-(k^x-eBy)^2-ieB\beta_s+S_s\right)h_s(y)=0
\label{dirac9M}
\end{eqnarray}
where $S_s$ is the separation coefficient. We can create four sets of pairs of equations, each corresponding to a different eigenvalue choice. For example, for $\lambda_1=1$, $\beta_1 =-i$ and $\lambda_2=1$, $\beta_2 =i$, we respectively have,
\begin{eqnarray}
\left(\partial_z^2+(k^0+qEz)^2-m^2-ieE-S_1\right)f^{(p)}_1(z)=0,\;\left(\partial_y^2-(k^x-eBy)^2-eB+S_1\right)h_1(y)=0 \nonumber\\ \left(\partial_z^2+(k^0+qEz)^2-m^2-ieE-S_2\right)f^{(p)}_2(z)=0,\; \left(\partial_y^2-(k^x-eBy)^2+eB+S_2\right)h_2(y)=0 \nonumber\\
\label{dirac9M'}
\end{eqnarray}
 Let us first focus on the $y$ differential equation, and likewise, \cite{HSSS}, it is the Hermite differential equation, with the separation constants,
\begin{equation}  
S_1=S_3=2(n_L+1)eB   \quad \text{and} \quad S_2=S_4=2n_LeB
\end{equation} 
 
where $n_L=0,1,2, \dots$ stands for the Landau level. Thus, we have the normalised solutions,
\begin{equation}
    \label{YTI}
    h_1(y)=h_2(y)=\left(\frac{\sqrt{|eB|}}{2^{n_L+1}\sqrt{\pi}(n_L+1)!}\right)^{1/2}e^{-y_+^2/2}H_{n_L}(y_-)=h_{n_L}(y_-)
\end{equation}
where $H_{n_L}(y_-)$ are the Hermite polynomials of order $n_L$ and $y_\pm=\sqrt{eB}(y\pm\frac{k^x}{eB})$.\\

Now, let us focus on the $z$-differential equation, for which we define the variable
$
	\zeta
=
	e^{i \pi/4}
	\sqrt{2|eE|}
	(
	z + {k^0}/{eE}
	)
$,
the solution of equation are $D_{\nu_s}(\pm \zeta)$ and $(D_{\nu_s}(\pm \zeta))^*=D_{-\nu_s-1}(\pm i\zeta)$. Also,
$
	\nu_s
=
	-(1+i\mu_s)/2
%=
%	-
%	\frac{1}{2}	
%	-
%	\frac{i}{2}	
%	\mu
$,
% $\kappa_{qE}$ is defined as $\kappa_{qE} = \pm 1$ for $qE\gtrless 0$, respectively.
with the parameter $\mu_s$ given by
\begin{eqnarray}
	\mu_s
=
	\frac{ m^2+S_s}{eE}
\label{muF}
% \qquad\qquad
%     \text{for the Landau levels $n_L=0,1,2,\cdots$.}
\end{eqnarray}
%Similar to the saclar field, we consider a particle that is incoming in the $z$-direction at $\absN{z}\to\infty$. %For the particle mode, the solution of \ref{2ndorder} is $F^{p}_1(x)\propto D_{\nu_1}(\zeta_+) h_1(y)$ and $F^{p}_2(x)\propto D_{-\nu_2-1}(i\zeta_+))^* h_2(y)$ for both the spins. Similarly, one can obtain the antiparticle modes by applying the charge conjugation operator $C=i\gamma^2$ on the complex conjugates of particle modes. However, 
Similar to the case of the scalar field, in this scenario as well, we can establish a distinction between the `in' and `out' sets of modes in a similar manner \cite{Ebadi:2014ufa, HSSS}.
Full Dirac `in' and `out' modes sets are defined by $\Psi(x)$ in \ref{DiracMinkowski} and  given as 
\begin{equation}
\label{modef1}
     U^{n_L, \text{in}}_{k,s}(x)  =\frac{1}{M_s}\left(i \gamma^{\mu} \partial_{\mu}-e \gamma^{\mu} A_{\mu}+m\right) e^{-i(k^0t-k^x x)}e^{-{y}_+^2/2}
	H_{n_L}({y}_-)
	D_{\nu_s} (\zeta)\epsilon_{s} \quad (s=1,2)
\end{equation}
\begin{equation}
\label{modef2}
V^{n_L, \text{in}}_{k,s}(x)  =\frac{1}{M_s}\left(i \gamma^{\mu} \partial_{\mu}-e \gamma^{\mu} A_{\mu}+m\right) e^{i(k^0t-k^x x)}e^{-{y}_-^2/2}
	H_{n_L}({y}_+)[D_{\nu_s} (\zeta)]^* \epsilon_{s} \quad (s=3,4)    
\end{equation}
\begin{equation}
\label{modef3}
 U^{n_L, \text{out}}_{k,s}(x)  =\frac{1}{M_s}\left(i \gamma^{\mu} \partial_{\mu}-e \gamma^{\mu} A_{\mu}+m\right) e^{-i(k^0t-k^x x)}e^{-{y}_+^2/2}
	H_{n_L}({y}_-) D_{\nu_s}(-\zeta)\epsilon_{s} \quad (s=1,2)
\end{equation}
\begin{equation}
\label{modef4}
   V^{n_L, \text{out}}_{k,s}(x)  =\frac{1}{M_s}\left(i \gamma^{\mu} \partial_{\mu}-e \gamma^{\mu} A_{\mu}+m\right) e^{i(k^0t-k^x x)}e^{-{y}_-^2/2}
	H_{n_L}({y}_+)
	[D_{\nu_s}(-\zeta)]^* \epsilon_{s} \quad (s=3,4) 
\end{equation}
Here, $U_{k,s}^{n_L}(x)$ and $V_{k,s}^{n_L}(x)$ represent modes for particles and antiparticles, respectively, with $M_s$ serving as the normalization factor. The orthogonality relations between these modes can be expressed in terms of the Dirac inner product as follows
\begin{eqnarray}
\begin{gathered}
\left(U_{k,s}^{n_L, \text { in }}, U_{p,r}^{n_L', \text { in }}\right)=\delta_{r s} \delta(k^0 - p^0)
    \delta(k^x - p^x)
    \delta_{n_L n_L'} \quad \left(U_{k,s}^{n_L, \text { out }}, U_{p,r}^{n_L', \text { out }}\right)=\delta_{r s} \delta(k^0 - p^0)
    \delta(k^x - p^x)
    \delta_{n_L n_L'}\\
\left(V_{k,s}^{n_L, \text { in }}, V_{p,r}^{n_L', \text { in }}\right)=\delta_{r s} \delta(k^0 - p^0)
    \delta(k^x - p^x)
    \delta_{n_L n_L'} \quad \left(V_{k,s}^{n_L \text { out }}, V_{p,r}^{n_L', \text { out }}\right)=\delta_{r s} \delta(k^0 - p^0)
    \delta(k^x - p^x)
    \delta_{n_L n_L'}
\end{gathered}
\end{eqnarray}

%Using the inner product, one can also find out the Bogoliubov coefficients defined as
%$$
%\begin{gathered}
%\left(U_{k}^{s,n_L, \text { in }}, U_{p}^{r,n_L, \text { out %}}\right)=\delta_{r s} \delta(\vec{k}-\vec{p}) \alpha_{k}^{s} \\
%\left(U_{k}^{s, n_L,\text { in }}, V{p}^{r, n_L,\text { out %}}\right)=\delta_{r s} \delta(\vec{k}-\vec{p}) \beta_{k}^{s},
%\end{gathered}
%$$
%with all the other inner products vanishing.
In terms of these orthonormal modes, the canonical field quantization is
\begin{eqnarray}
\psi(x) &&= \sum_{n_L; s=1,2}\int\frac{dk^0 dk^x}{2\pi} \Bigg[a^{\rm in}_{k,s,n_L}U_{k,s}^{n_L, \text { in }}+b^{\dagger \rm in}_{k,s,n_L}V_{k,s}^{n_L, \text { in }}\Bigg] \nonumber\\
&&=\sum_{n_L; s=1,2}\int\frac{dk^0 dk^x}{2\pi }\Bigg[a^{\rm out}_{k,s,n_L}U_{k,s}^{n_L, \text { out }}+b^{\dagger \rm out}_{k,s,n_L}V_{k,s}^{n_L, \text { out }}\Bigg]
\label{fieldMF}
\end{eqnarray}
These creation and annihilation operators satisfy the usual canonical anti-commutation relations. These modes are related via the Bogoliubov transformation as
\begin{equation}
	U_{k,s}^{n_L, \text { in }}
=
	\alpha_{k,s}
	U_{k,s}^{n_L, \text { out }}
	+
	\beta_{k,s}
	\left(
	V_{-k,s}^{n_L, \text { out }}
	\right)^*
\label{eq:Bogoliuboftrf0F}
\end{equation}
where $\alpha_{k,s}$ and $\beta_{k,s}$ are the Bogoliubov coefficients given as
%. Using some properties of the parabolic cylinder functions, we obtain the Bogoliubov coefficients given as
\begin{eqnarray}
 \alpha_{k,s}=\sqrt{\frac{\mu_s}{\pi}} \Gamma\left(\frac{i \mu_s}{2}\right) \sinh \left(\frac{\pi \mu_s}{2}\right) e^{-\frac{\pi \mu_s}{4}}, \quad \quad
 \beta_{k,s}=e^{- \frac{\pi\mu_s}{2}}    
\end{eqnarray}
These coefficients are obtained by taking the following inner products \ref{eq:Bogoliuboftrf0F}, \be
(U^{n_L,\text{in}}_{k,s},U^{n_L,\text{out}}_{p,r})=\alpha_{k,s} \delta_{rs}\delta(k-p) \;\text{and} \;(U^{n_L,\text{in}}_{k,s},V_{-p,r}^{n_L, \text { out }})=\beta_{k,s} \delta_{rs}\delta(k-p)\ee
These coefficients satisfy the consistency relation $|\alpha_{k,s}|^{2}+|\beta_{k,s}|^{2}=1$. It is worth noting that there exists a sign change in the Bogoliubov consistency condition compared to scalars \cite{HSSS}. This distinction arises from the fact that fermions adhere to an anticommutation relation, while scalars follow a commutation relation. Notably, when examining the expression for $\mu_s$, which resembles that of scalar field theory, an interesting behaviour emerges. As the electric field parameter $E$ approaches zero, $\beta_s$ also approaches zero, regardless of the strength of the external magnetic field $B$. Additionally, when keeping $E$ fixed, $\beta_s$ exhibits a monotonic decrease as $B$ increases.

%This behaviour suggests intriguing insights into the system's physics, particularly in connection to the ground state or vacuum state and the impact of an increasing external magnetic field on particle creation or annihilation processes described by the Bogoliubov coefficients.

Next, on employing the orthonormality conditions between the modes, we derive the relationship between the `in' and the `out' operators as 
\begin{equation}
\label{in_outMF}
 a^{\rm \text{out}}_{k,s,n_L}\;=\;\alpha_{k,s}a^{\rm \text{in}}_{k,s,n_L}-\beta_{k,s}^{*} b^{\dagger \rm \text{in}}_{k,s,n_L},\quad
 b^{\rm \text{out}}_{k,s,n_L}\;=\;\alpha_{k,s} a^{\text{in}}_{k,s,n_L}+\beta_{k,s}^{*} a^{\dagger \rm \text{in}}_{k,s,n_L}
\end{equation}
Subject to the field quantisation in \ref{fieldMF}, the `in' and the `out' vacuum are defined as,
\begin{equation}
  a^{ \rm \text{in}}_{k,s,n_L}|0\rangle_{\rm \text{in}}=0 = b^{ \rm \text{in}}_{k,s,n_L} |0\rangle_{\rm in}\qquad {\rm and }\qquad a^{ \rm \text{out}}_{k,s,n_L}|0\rangle_{\rm \text{out}}=0 = b^{ \rm \text{out}}_{k,s,n_L}|0\rangle_{\rm \text{out}}
\end{equation}
Thus, the Bogoliubov relations  \ref{in_outMF} imply a squeezed state expansion between the `in' and `out' states for a given momentum, which reads as
\begin{equation}
\label{vaccum1MF}
|0\rangle_{\rm in}\;
= \left(\alpha_{k,1} |0_{k}^{(1)}\,0_{-k}^{(1)}\rangle_{\rm out} +\beta_{k,1} |1_{k}^{(1)}\,1_{-k}^{(1)}\rangle_{\rm out}\right )\otimes  \left(  \alpha_{k,2} |0_{k}^{(2)}\,0_{-k}^{(2)}\rangle_{\rm out} +\beta_{k,2} |1_{k}^{(2)}\,1_{-k}^{(2)}\rangle_{\rm out}\right)
\end{equation}
When considering a large range of values for the parameter $\mu_s$, it becomes equivalent to the initial vacuum state. To express the excited states labelled as `in' in terms of the `out' states, we utilize the `in' creation operators on the left side of equation \ref{vaccum1MF} and apply the Bogoliubov relations defined in equation \ref{in_outMF}. Finally, we note that $s=1\; \text{and}\; s=2$ sectors are factorised in \ref{vaccum1MF}, leading to 
\be |0\rangle_{\rm in}  = |0^{(1)}\rangle_{\rm in}\otimes |0^{(2)}\rangle_{\rm in}\ee
For the sake of simplicity, we can focus solely on one sector, let's say $|0^{(1)}\rangle_{\rm in}$, of the in-vacuum. In this context, $\mu_1$ corresponds to the parameter $\mu$ in the scalar field case \cite{HSSS}. From this point forward, we will use $\mu$ instead of $\mu_1$. Note that in all the figures, an increase in $\mu$ corresponds to an increase in the magnetic field ($B$), mass ($m$), and Landau level ($n_L$), along with a decrease in the electric field. Conversely, a decrease in $\mu$ corresponds to an increase in the electric field and a decrease in the magnetic field ($B$), mass ($m$), and Landau level ($n_L$).

%It is essential to note that the particle density in the `in' vacuum state matches that of the complex scalar field.
%%%%%%%%%%%%%%%%%%%%%%%%%%%%%%%%%%%%%%

\section{Correlation measures}
\label{Correlation measures}
The three fundamental questions that arise naturally are: ``Is a state correlated (or entangled) or not?"; ``How does the level of correlation in this state compare to that in another state?"; and "How can these correlations be quantified?". Various measures have been developed to answer these questions, and they have been extensively studied across a broad range of theoretical research, as discussed in \cite{Plenio:2007zz, Zyczkowski:1998yd, Martin, Vidal:2002zz, wang, Plenio:2005, Calabrese:2012nk, Nishioka:2018khk, jmath} and references therein.

\subsection{Entanglement entropy}
Consider a bipartite system consisting of subsystems \( A \) and \( B \), where the Hilbert space is decomposed as \( H_{AB} = H_A \otimes H_B \). Let \( \rho_{AB} \) represent the density matrix of states on \( H_{AB} \). In such a system, the entanglement entropy of subsystem \( A \) is defined by the von Neumann entropy of its reduced density matrix \( \rho_A \). It is specifically given by:

\be
S(\rho_A) = -\text{Tr}_A \left( \rho_A \ln \rho_A \right),
\ee

where \( \rho_A \) is obtained by tracing out subsystem \( B \) from the joint density matrix \( \rho_{AB} \):

\be
\rho_A = \text{Tr}_B \left( \rho_{AB} \right).
\ee

Similarly, the entanglement entropy of subsystem \( B \) is defined analogously. For a pure state of the total system, the entanglement entropies of subsystems \( A \) and \( B \) are equal:

\be
S(\rho_A) = S(\rho_B).
\ee

Moreover, if the joint state \( \rho_{AB} \) is separable (i.e., \( \rho_{AB} = \rho_A \otimes \rho_B \)), the entanglement entropy vanishes:

\be
S(\rho_A) = S(\rho_B) = 0.
\ee

The von Neumann entropies also satisfy the subadditivity inequality:

\be
S(\rho_{AB}) \leq S(\rho_A) + S(\rho_B),
\ee

where \( S(\rho_{AB}) \) is the von Neumann entropy of the joint state \( \rho_{AB} \). Equality holds if and only if \( \rho_{AB} \) is separable.

\subsection{Bell's inequality violation} 
\label{Bell's inequality violation}
The construction of Bell or Bell-Mermin-Klyshko (BMK) operators for fermions is analogous to that for scalar fields, as detailed in \cite{ bell:2017, NielsenChuang} and related references. Consider two pairs of non-commuting observables defined over the Hilbert spaces \( \mathcal{H}_{A} \) and \( \mathcal{H}_{B} \), namely \( (\mathit{O}_{1}, \mathit{O}'_{1}) \in \mathcal{H}_{A} \) and \( (\mathit{O}_{2}, \mathit{O}'_{2}) \in \mathcal{H}_{B} \). We assume these are spin-\(\frac{1}{2}\) operators oriented along specific directions, such as \( \mathit{O} = n_{i} \sigma_{i} \) and \( \mathit{O}' = n_{i}^{\prime} \sigma_{i} \), where \( \sigma_{i} \) are the Pauli matrices and \( n_{i} \), \( n_{i}^{\prime} \) are unit vectors in three-dimensional Euclidean space. The eigenvalues of each of these operators are \( \pm 1 \). The Bell operator, \( \mathcal{B} \in \mathcal{H}_{A} \otimes \mathcal{H}_{B} \), is defined as (with the tensor product sign suppressed):

\begin{equation}
    \label{op2}
\mathcal{B} = \mathit{O}_{1}(\mathit{O}_{2} + \mathit{O}'_{2}) + \mathit{O}'_{1}(\mathit{O}_{2} - \mathit{O}'_{2}).
\end{equation}

In classical local hidden variable theories, Bell's inequality states that:

\be
\langle \mathcal{B}^{2} \rangle \leq 4 \quad \text{and} \quad |\langle \mathcal{B} \rangle| \leq 2.
\ee

This inequality is violated in quantum mechanics, as can be demonstrated by calculating $\mathcal{B}^{2}$ by using the expression $\mathcal{B}$ from \ref{op2}:

\begin{equation}
\mathcal{B}^{2} = \mathbf{I} - [\mathit{O}_{1}, \mathit{O}'_{1}][\mathit{O}_{2}, \mathit{O}'_{2}],
\end{equation}
where \( \mathbf{I} \) is the identity operator. Utilizing the commutation relations of the Pauli matrices, we obtain

\be |\langle \mathcal{B} \rangle| \leq 2 \sqrt{2}, \ee 
which demonstrates a violation of Bell's inequality, with equality corresponding to the maximum possible violation. For a detailed study, we refer the reader to \cite{NielsenChuang}. This construction can be extended to multipartite systems with pure density matrices, such as squeezed states formed by mixing different modes. For more details, we refer the reader to \cite{bell:2017} and related references.

%We wish to investigate Bell's inequality violation for the vacuum as well as some maximally entangled initial states. 

\subsection{Quantum mutual information}
\label{sec:QMI}
%%%%%%%%%
\noindent

The quantum mutual information quantifies both quantum and classical correlations between the subsystems \( A \) and \( B \). For the state \( \rho_{AB} \), it is defined as:

\be
I(A,B) = S(\rho_A) + S(\rho_B) - S(\rho_{AB}).
\ee

The mutual information is always non-negative, \( I(A,B) \geq 0 \), which follows directly from the subadditivity of the entanglement entropy. Equality, \( I(A,B) = 0 \), holds if and only if the joint state \( \rho_{AB} \) is separable, i.e., \( \rho_{AB} = \rho_A \otimes \rho_B \). Further properties of the quantum mutual information can be found in sources such as \cite{NielsenChuang}.

\section{Number density, entanglement entropy and Bell's inequality violation of the fermionic field vacuum}
\label{ND}

We wish to compute number density, entanglement entropy, and Bell's inequality violation for the vacuum state $|0^{(1)}\rangle_{\rm in}$ in the presence of constant strength electric and magnetic fields.% Thus for simplicity, we can work only with a single sector, say $|0_{k}^{(1)}\rangle^{\rm \text{in}}$, of the in-vacuum. 

\subsection{Number density}
The number density of particles and antiparticles in $|0^{(1)}\rangle_{\rm in}$ is given as
\begin{equation}
    \label{Nd}
_{\text{in}}\langle0^{(1)}| a_{k,s,n_L}^{ \text{out}}a_{k,s,n_L}^{\dagger \text{out}}|0^{(1)}\rangle_{\text{in}}=_{\text{in}}\langle0^{(1)}| b_{k,s,n_L}^{ \text{out}}b_{k,s,n_L}^{\dagger \text{out}}|0^{(1)}\rangle_{\text{in}}=|\beta|^2=e^{-\pi\mu}
\end{equation}
it represents the density of created particles. The dependence of \( |\beta|^2 \) on \( \mu \) is shown in \ref{fig:FNDM} (a), where 
\begin{equation}
\label{eq:munew}
\mu = \frac{m^2 + 2(n_L+1)eB}{eE} 
\end{equation}

Here, an increase in \( \mu \) corresponds to an increase in \( B \) and a decrease in \( E \), with \( m \) and \( n_L \) held constant. As \( \mu \) decreases, \( |\beta|^2 \) increases, reaching its maximum value of \( |\beta|^2 = 1 \) in the limit \( \mu \to 0 \). Conversely, \( |\beta|^2 \to 0 \) as \( \mu \to \infty \), indicating that the vacuum reverts to the initial pure state. This behavior reflects the suppression of pair creation due to the stabilization of the vacuum as \( B \) increases. The underlying mechanism is that the magnetic field exerts a Lorentz force \( q(\mathbf{v} \times \mathbf{B}) \) on the particles and antiparticles created by the electric field, which move in opposite directions. This force acts in the same direction for both particles, thereby suppressing their separation and reducing pair production.

\begin{comment}

This case is the same as that of the scalar case \cite{HSSS}. The variation of the number density \ref{Nd} with respect to the parameter $\mu$ is shown in the left plot of \ref{fig:FNDM}. It decreases with the increase in the strength of the electric field (which corresponds to the increasing strength of $\mu$) and increases with the increase in the strength of the magnetic field (which corresponds to the decreasing strength of $\mu$).
\end{comment}

\subsection{Entanglement entropy}
Next, we wish to compute the entanglement between the created particles and antiparticles in the vacuum state. The density matrix corresponding to this state is pure, $\rho= |0^{(1)}\rangle_{\rm \text{in}\,\,\text{in}}\langle0^{(1)}|$. Using \ref{vaccum1MF}, we write down $\rho$ in terms of the out states, which contain both $k$ and $-k$ degrees of freedom. The reduced density matrix corresponding to the $k$ sector (say, particle) is given by, 
\be
	\rho_k
=
	\text{Tr}_{-k}
	(\rho)
=
	|\alpha|^{2} |0_{k}^{(1)}\rangle_{\rm \text{out}\,\, \text{out}}\langle 0_{k}^{(1)}| + |\beta|^{2} |1_{k}^{(1)}\rangle_{\rm \text{out}\,\, \text{out}}\langle 1_{k}^{(1)}|
\ee and hence the entanglement entropy
is given by 
\begin{equation}
    \label{entropyMF}
	S^F_k =
	-
	\text{Tr}_k
	\rho_k
	\ln \rho_k
=
	-   \left[ \ln	(1-|\beta|^{2})    +|\beta|^2  \ln \frac{|\beta|^{2}}{1-|\beta|^2}\right]
\end{equation}
\begin{comment}
    
\begin{figure}[h]
    \centering
    \includegraphics[scale=.47]{figures/FermionND.pdf}
    \includegraphics[scale=.84]{figures/EEFM.pdf}
    \includegraphics[scale=.60]{figures/BVFM.pdf}
    \caption{\it {\small The variation of number density, entanglement entropy, and Bell's inequality violation of the fermionic vacuum state with respect to the parameter $\mu(=\frac{ m^2+2(n_L+1)eB}{eE})$. \textcolor{red}{For very small or very large electric fields, the number density approaches $0$ and $1$, respectively. Similarly, the maximum value of Bell's operator reaches $2$ under these extreme conditions, with its peak occurring at $\mu =  \log{(2)}/\pi$. Likewise, the entropy reaches its minimum value of $0$ and its maximum value of $1$ at $\mu =  \log{(2)}/\pi$. In contrast, the behavior of the magnetic field is opposite to that of the electric field, influencing the system in an inverse manner.}}}
    \label{fig:FNDM}
\end{figure}

\end{comment}

\begin{figure}[h]
    \centering
    % Subfigure (a)
    \begin{subfigure}[b]{0.32\textwidth}
        \centering
        \includegraphics[scale=.47]{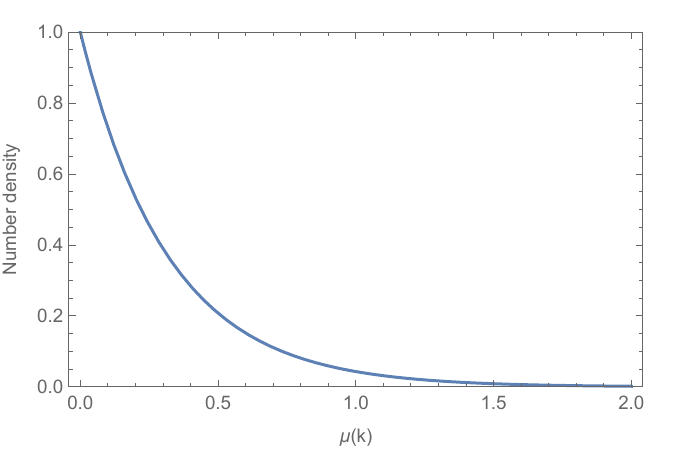}
        \caption{Number density}
    \end{subfigure}
    % Subfigure (b)
    \begin{subfigure}[b]{0.32\textwidth}
        \centering
        \includegraphics[scale=.84]{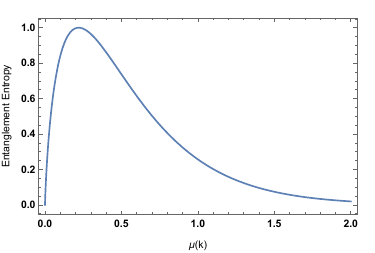}
        \caption{Entanglement entropy}
    \end{subfigure}
    % Subfigure (c)
    \begin{subfigure}[b]{0.32\textwidth}
        \centering
        \includegraphics[scale=.60]{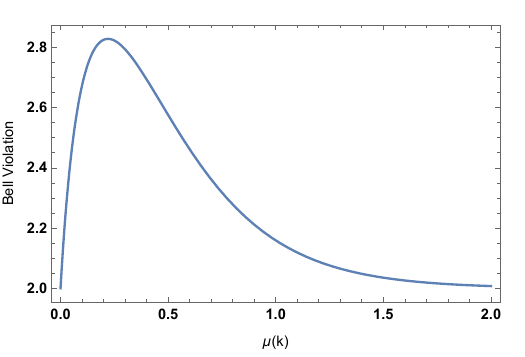}
        \caption{Bell's inequality violation}
    \end{subfigure}

    \caption{\it {\small The variation of number density, entanglement entropy, and Bell's inequality violation of the fermionic vacuum state with respect to the parameter $\mu(=\frac{ m^2+2(n_L+1)eB}{eE})$. For very small or very large electric fields, the number density approaches $0$ and $1$, respectively. Similarly, the maximum value of Bell's operator reaches $2$ under these extreme conditions, with its peak occurring at $\mu = \log{(2)}/\pi$. Likewise, the entanglement entropy reaches its minimum value of $0$ and its maximum value of $1$ at $\mu = \log{(2)}/\pi$. In contrast, the behavior of the magnetic field is opposite to that of the electric field, influencing the system in an inverse manner.}}
    \label{fig:FNDM}
\end{figure}

\begin{comment}
\begin{figure}[h]
    \centering
    \includegraphics[scale=.70]{figures/BVFM.pdf}
    \caption{\it{\small The variation of the Bell violation of the fermionic vacuum state with respect to the parameter $\mu$.}}
    \label{fig:FNDM}
\end{figure}
\end{comment}

The variation of \ref{entropyMF} with respect to parameter $\mu$ is shown in the middle of \ref{fig:FNDM}. Unlike the complex scalar field, it behaves in a non-monotonic manner. Firstly, it increases and reaches its maximum value and then decreases with the increase in the parameter $\mu$. As the electric field strength grows, the entanglement entropy exhibits an initial rise, followed by a decline after reaching its peak. This behaviour concurs with the findings reported in \cite{Ebadi:2014ufa}. Additionally, in contrast to the complex scalar field, the entanglement entropy for the fermionic field, as discussed in \cite{HSSS}, displays a non-monotonic pattern and does not approach zero at high $\mu$ values; it maintains a non-zero value in that limit.%Such non-monotonic behaviour of the entanglement entropy of the fermionic field in the presence of a background electric field was observed earlier by \cite{Ebadi:2014ufa}.
% \begin{figure}
%    \centering
%    \includegraphics[scale=0.6]{figures/FermionEE.pdf}
%    \caption{The variation of the entanglement entropy of the fermionic field vacuum (\ref{entropyMF}) with respect to parameter $\mu$.}
%     \label{fig:EEMF}
% \end{figure}

\subsection{Bell's inequality violation}
We wish to find out the expectation value of $\mathcal{B}$, with respect to the vacuum state $|0_{k}^{(1)} \rangle^{\rm in}$, given at the end of \ref{Fermionic field in background electric and magnetic field}. In order to do this, one usually introduces pseudospin operators measuring the parity in the Hilbert space along different axes, e.g~\cite{bell:2017} and references therein. These operators for fermionic systems with eigenvalues $\pm 1$ are defined as,
\begin{equation}
\label{unitvector}
\mathbf{\hat{n}}.\mathbf{S} = S_{z}\cos\theta + \sin\theta(e^{i\phi}S_{-}+e^{-i\phi}S_{+}), 
\end{equation}
where $\mathbf{\hat{n}} = (\sin\theta \cos\phi,\sin\theta \sin\phi,\cos\phi)$ is a unit vector in the Euclidean $3$-plane. The action of the operators $S_z$ and $S_{\pm}$ are defined on the {\it out states},
\begin{equation}
\label{spin}
S_z | 0\rangle = - | 0\rangle, \quad S_z | 1\rangle =  | 1\rangle, \quad S_+| 0\rangle =| 1\rangle, \quad S_+| 1\rangle =0, \quad S_-| 0\rangle =0, \quad S_-| 1\rangle =|0\rangle
\end{equation}
Without any loss of generality, we take the operators to be confined to the $x-z$ plane, so that we may set  $\phi = 0 $ in \ref{unitvector}. We may then take $\mathit{O_{i}}=\;\mathbf{\hat{n}}_i\cdot\mathbf{S}$  and $\mathit{O}'_i=\mathbf{\hat{n}'}_i\cdot\mathbf{S}$ 
with $i=1,2$. Here $\mathbf{\hat{n}}_i$ and $\mathbf{\hat{n}'}_i$ are two pairs of unit vectors in the Euclidean $3$-plane, characterised by their angles with the $z$-axis, $\theta_{i}$, $\theta'_{i}$ (with $i=1,2$) respectively. 

Using the above constructions, and the squeezed state expansion \ref{vaccum1MF}  and also the operations \ref{spin} defined on the out states,  the desired expectation value is given by,
\begin{equation}
\label{B2}
^{\rm in}\langle 0^{(1)}_{k}|\mathcal{B}|0_{k}^{(1)}\rangle^{\rm in} = [E(\theta_{1},\theta_{2})+E(\theta_{1},\theta'_{2})+E(\theta'_{1},\theta_{2})-E(\theta'_{1},\theta'_{2})]
\end{equation}
where, $\mathit{O}_i$ and $\mathit{O}'_i$ are assumed to operate respectively on the $k$ and $-k$ sectors of the out states in \ref{vaccum1MF}, and 
\be E(\theta_{1},\theta_{2})
=
\cos\theta_{1} \cos\theta_{2} + 2|\alpha\beta| \sin\theta_{1} \sin\theta_{2}\ee
Choosing now s, we have from \ref{B2},
\begin{equation}
^{\rm in}\langle 0_{k}^{(1)}|\mathcal{B}|0_{k}^{(1)}\rangle^{\rm in} = 2(\cos\theta_{2}+ 2|\alpha\beta|\sin\theta_{2})
\label{bvpm}
\end{equation}
The above expression maximises at $\theta_{2} =\tan^{-1}(2|\alpha\beta|)$, so that the above expectation value becomes
\be \langle\mathcal{B}\rangle_{\rm max} = 2\left(1+ 4|\alpha \beta|^2 \right)^{1/2}\ee
Thus $\langle\mathcal{B}\rangle_{\rm max}\geq2$, and hence there is  Bell violation for $|\beta|>0$.
We have plotted $\langle\mathcal{B}\rangle_{\rm max}$ in~\ref{fig:FNDM} (right most) with respect to the parameter $\mu$. As of the vacuum entanglement entropy, \ref{fig:FNDM} (middle), the Bell violation firstly increases with the increase in $\mu$ and after reaching its maxima it decreases monotonically with the increasing $\mu$ and reaches the value two. Once again, this happens due to the suppression of particle creation by the magnetic field.

Note that the vacuum state is pure. Instead of a vacuum, if we consider a pure but maximally entangled state, make its squeezed state expansion, and then trace out some parts of it in order to construct a bipartite subsystem, the resulting density matrix turns out to be mixed. The above construction is valid for pure ensembles only and one requires a different formalism to deal with mixed ensembles, e.g.~\cite{bell_3}. We wish to study such cases below, to demonstrate their qualitative differences with the vacuum case.

\section{Maximally entangled state}
\label{Maximally entangled state}
In this section, we have focused on a maximally entangled state constructed using two fermionic fields.  For computational simplicity, we assume that both the fields have the same rest mass, and we consider modes in which their momenta along the $z$-direction  and the Landau levels are the same. We have quantified the correlations between various sectors of these states using measures such as mutual information and Bell's inequality violation.

We begin by considering the initial state,
\begin{equation}
\label{Bell-1MF}
|\psi\rangle=\frac{|0_p 0_{-p}0_k0_{-k}\rangle^{\text{in}}+|0_p1_{-p}1_{k}0_{-k}\rangle^{\text{in}}}{\sqrt{2}}
\end{equation}
has zero net charge. %In the four entries of a ket above, the first pair of states corresponds to one fermionic field, whereas the last pair corresponds to another. The $\pm$-sign in front of the momenta stands respectively for the particle and anti-particle degrees of freedom.
Using then \ref{in_outMF} and \ref{vaccum1MF}, we re-express \ref{Bell-1MF} in the out basis as
\begin{eqnarray}
\label{p}
|\psi\rangle=\frac{(\alpha|0_p0_{-p}\rangle^{\rm out}+\beta|1_p1_{-p}\rangle^{\rm out})(\alpha|0_k0_{-k}\rangle^{\rm out}+\beta|1_k1_{-k}\rangle^{\rm out})+|0_p1_{-p}\rangle^{\rm out}|1_{k}0_{-k}\rangle^{\rm out}}{\sqrt{2}}
\end{eqnarray}
at the large $\mu$ range, this state coincides with the initial state.

We focus on the correlations between the particle-particle (or antiparticle-antiparticle), particle-antiparticle and antiparticle-particle sectors corresponding to the density matrix of the above state. Accordingly, tracing out first the antiparticle-antiparticle degrees of freedom of the density matrix $\rho^{(0)}\;=\;|\psi\rangle \langle \psi|$, we construct the reduced density matrix for the particle-particle given as 
\begin{eqnarray}
\rho_{p,k}^{(0)}=\rho_{-p,-k}^{(0)}=\;\frac{1}{2}\left(\begin{array}{cccc}
|\alpha|^{4}&0&0&0\\
0&|\alpha\beta|^2&(\alpha\beta)^* &0\\
0&\alpha\beta&|\alpha\beta|^2+1&0\\
0&0&0&|\beta|^{4}\\
\end{array}\right)
\label{pp}
\end{eqnarray}
We note that the reduced density matrix of the antiparticle-antiparticle $(\rho_{-p,-k}^{(0)})$ sector is the same as the particle-particle sector $(\rho_{p,k}^{(0)})$. This equivalence arises because the quantum state $|\psi\rangle$ \ref{p} is symmetric under the interchange of $p$, $k$ with $-p$, $-k$ and the process of tracing respects this symmetry. The symmetry implies that the correlations between the modes $p$, $k$ are identical to those between the modes $-p$, $-k$.
Likewise, we obtain the reduced density matrix for particle-antiparticle and antiparticle-particle sectors,
\begin{equation}
\rho_{-p,k}^{(0)}\;= \frac{1}{2}\left( {\begin{array}{cccc}
|\alpha|^{4}&0&0&(\alpha^{*})^2\\
0&|\alpha\beta|^2&0&0\\
0&0&|\alpha \beta|^2&0\\
\alpha^2&0&0&|\beta|^{4}+1\\
\end{array}}\right)
\label{ap}
\end{equation}
\begin{equation}
\rho_{p,-k}^{(0)}\;= \frac{1}{2}\left( {\begin{array}{cccc}
|\alpha|^{4}+1&0&0&\beta^2\\
0&|\alpha\beta|^2&0&0\\
0&0&|\alpha \beta|^2&0\\
(\beta^{*})^2&0&0&|\beta|^{4}\\
\end{array}}\right)
\label{pa}
\end{equation}
The reduced density matrix for the antiparticle-particle sector $(\rho_{-p,k}^{(0)})$, is the same as $\rho_{p,-k}^{(0)}$. Next, we have computed Bell's inequality violation and mutual information of these sectors. 

%The logarithmic negativity of $\rho^{(0)}_{kp}$  is

\begin{comment}   
The variation of the logarithmic negativity of $\rho_{kp}^{0}$ (\ref{LNrhokp}) with respect to the parameter $\mu$ is shown in \ref{fig:MIMF}. It decreases monotonically with the increase in $\mu$, whereas it is zero for $\rho^{(0)}_{-k,p}$; however, zero logarithmic negativity does not correspond to no entanglement state.
\end{comment}

\subsection{Bell's Inequality violation}

Since these reduced-density matrices are mixed. For such a system, a procedure to compute the Bell inequality has been proposed  in~\cite{seprability}. According to this, any density matrix ($\rho$) in a Hilbert-Schmidt basis can be decomposed as
\begin{equation}
    \label{decomrho}
\rho=\frac{1}{4}\Big(I\otimes I+\mathit{O_i}\otimes I+I\otimes \mathit{O}'_i+\sum_{i,j=1}^{3}T_{ij}\sigma_i \otimes \sigma_j\Big)
\end{equation}
where $I$ stands for the identity operator and $\sigma_i$'s are the Pauli matrices. Here, $\mathit{O_{i}}=\;\mathbf{\hat{n}}_i\cdot\mathbf{S}$  and $\mathit{O}'_i=\mathbf{\hat{n}'}_i\cdot\mathbf{S}$  are the operators along specific directions in the $3$-dimensional Euclidean space as defined earlier in \ref{Bell's inequality violation}. $T$ is called the correlation matrix for the general decomposition of $\rho$, \ref{decomrho}, and it can be easily checked that its elements are given by  $T_{ij} \equiv {\rm Tr}[\rho \sigma_{i}\otimes \sigma_{j}]$. In the decomposition \ref{decomrho}, the middle two terms describe the local behaviour of the state, and the last term describes the correlations of the state. 

We wish to compute the maximum average value of the Bell operator defined in \ref{op2} with respect to the density matrix $\rho$, \ref{decomrho}. Following \cite{bellmix}, one can choose
\begin{equation}
    \left(\mathbf{\hat{n}}_2+\mathbf{\hat{n}'}_2\right)= 2 \cos{\theta} \mathbf{\hat{e}},\qquad \qquad\left(\mathbf{\hat{n}}_2-\mathbf{\hat{n}'}_2\right)= 2 \cos{\theta} \mathbf{\hat{e}'}
\end{equation}
where $\mathbf{\hat{e}}$ and $\mathbf{\hat{e}'}$ are a pair of orthonormal vectors, and $\theta \in [0,\pi/2]$. Next, we choose $\mathbf{\hat{n}}_1$ and $\mathbf{\hat{n}'}_1$ in the direction of $T\mathbf{\hat{e}}$ and $T\mathbf{\hat{e}'}$, respectively. On substituting all these choices of unit vectors in \ref{op2}, the expectation value of the Bell operator becomes
\begin{equation}
    \label{Btheta}
   \langle \mathcal{B}\rangle=2(||T\mathbf{\hat{e}}||\cos \theta+||T\mathbf{\hat{e}'}||\sin \theta)
\end{equation}
where $||\;||$ denotes the Euclidean norm in $3-$dimensional space and defined as $||T\mathbf{\hat{e}}||=(\mathbf{\hat{e}},\;T^T T\mathbf{\hat{e}})$, where the round brackets correspond to the Euclidean scalar product and $T^T$ stands for transpose of $T$. On maximising \ref{Btheta} with respect to $\theta$, we have
\begin{equation}
    \label{Btheta1}
\langle \mathcal{B}\rangle=2\sqrt{||T\mathbf{\hat{e}}||^2+||T\mathbf{\hat{e}'}||^2}
\end{equation}
Let us now maximize the above average with respect to the choices of the unit vectors, $\mathbf{\hat{e}}$ and $\mathbf{\hat{e}'}$. We choose them as the eigenvectors corresponding to the maximum eigenvalues of the matrix $U$, defined as
\be U=(T)^{\rm T}T\ee
which is a symmetric matrix and can be diagonalised.  Having all these parameters fixed, thus the maximum value of the Bell operator is given by
\begin{equation}
    \langle \mathcal{B}_{\rm \text{max}}\rangle\;=\; 2\sqrt{\lambda_{1}+\lambda_{2}},
    \label{bv}
\end{equation}
where $\lambda_{1}$ and $\lambda_{2}$ are the two largest eigenvalues of the $(3\times 3)$ matrix $U$. Here also, the violation of the Bell inequality as earlier will correspond to $ \langle \mathcal{B}_{\rm max}\rangle>2$ in \ref{bv}. \\

\begin{figure}
    \centering
    \includegraphics[width=0.50\linewidth]{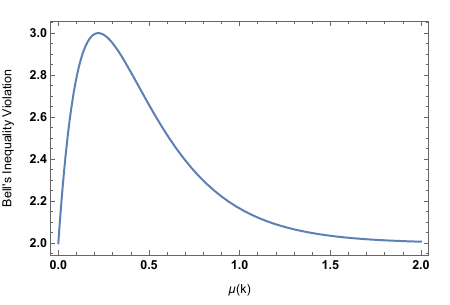}
    \caption{\it {\small The Bell's inequality violation of $\rho^{(0)}_{-p,k}$ ($\langle \mathcal{B}_{\rm max}\rangle (\rho^{(0)}_{-p,k})$) with respect to the parameter $\mu(=\frac{ m^2+2(n_L+1)eB}{eE})$. For extremely small or large electric fields, the Bell's inequality violation is approximately $2$. It attains the maximum value at $\mu = \log{(2)}/\pi$. In contrast, the
behavior of the magnetic field is opposite to that of the electric field, influencing the system in an inverse
manner.}}
    \label{FigB0apk}
\end{figure}

Next, we computed the maximum value of the Bell's operator for \ref{pp}, \ref{ap} and \ref{pa} and they are given as
\begin{equation}
    \langle \mathcal{B}_{\rm max}\rangle (\rho^{(0)}_{p,k})=\langle \mathcal{B}_{\rm max}\rangle (\rho^{(0)}_{-p,-k})=4|\alpha|^2|\beta|^2
    \label{Bpp}
\end{equation}
\begin{equation}
    \langle \mathcal{B}_{\rm max}\rangle (\rho^{(0)}_{-p,k})=2(1+|\alpha|^2|\beta|^2)
    \label{Bap}
\end{equation}
\begin{equation}
    \langle \mathcal{B}_{\rm max}\rangle (\rho^{(0)}_{p,-k})=2(1-|\alpha|^2|\beta|^2)
    \label{Bpa}
\end{equation}
 %We observe that only the antiparticle-particle sector ($ \langle \mathcal{B}_{\rm max}\rangle (\rho^{(0)}_{-p,k})$) shows Bell's inequality violation. However, the absence of a violation of Bell's inequality does not imply the absence of entanglement. Its variation with respect to the parameter $\mu$ is given by \ref{FigB0apk}. It is non-monotonic with respect to the parameter, $\mu$ and its behaviour is similar to that of the vacuum state.
We observe that Bell's inequality is violated only in the antiparticle-particle sector ($ \langle \mathcal{B}_{\rm max}\rangle (\rho^{(0)}_{-p,k})$). However, the absence of such a violation does not necessarily indicate the absence of entanglement. The variation of Bell's inequality violation with respect to the parameter $\mu$, as shown in \ref{FigB0apk}, is non-monotonic and exhibits a behavior similar to that of the vacuum state.

The violation of Bell's inequality in the antiparticle-particle sector, while absent in the particle-antiparticle and particle-particle (or antiparticle-antiparticle) sectors, can be attributed to the constraints imposed by entanglement monogamy. This principle ensures that strong quantum correlations in one sector diminish the possibility of equally strong correlations in other sectors, underscoring the inherently interdependent nature of quantum subsystems \cite{monogamy1, monogamy2, monogamy3}.

\subsection{Mutual Information}
The mutual information for the particle-particle $(\rho^{(0)}_{p,k})$, particle-antiparticle $(\rho_{p,-k}^{(0)})$ and antiparticle-particle $(\rho_{-p,k}^{(0)})$ sectors are as follows:
%\begin{equation}
 %\label{mi0kp}
%\begin{split}  
%I(\rho_{k,p}^{(0)})
%I=\frac{| \alpha | ^4}{2} \ln \frac{| \alpha | ^4}{2}+\Bigg(\frac{2 | \alpha | ^2 | \beta | ^2-\sqrt{4 | \alpha | ^2 | \beta | ^2+1}+1}{4} \Bigg) \ln\Bigg( \frac{2 | \alpha | ^2 | \beta | ^2-\sqrt{4 | \alpha | ^2 | \beta | ^2+1}+1}{4}\Bigg) \\-| \alpha | ^2 \ln \frac{| \alpha | ^2}{2}+\Bigg(\frac{2 | \alpha | ^2 | \beta | ^2 +\sqrt{4 | \alpha | ^2 | \beta | ^2+1}+1}{4} \Bigg) \ln \Bigg(\frac{2 | \alpha | ^2 | \beta | ^2+\sqrt{4 | \alpha | ^2 | \beta | ^2+1}+1}{4}\Bigg) \\+\frac{| \beta | ^4}{2}  \ln \frac{| \beta | ^4}{2}-| \beta | ^2 \ln \frac{| \beta | ^2}{2}
%\end{split}
%\end{equation}
\begin{equation}
 \label{mi0kp}
I(\rho^{(0)}_{p,k})=\frac{| \alpha | ^4}{2} \ln \frac{| \alpha | ^4}{2}-| \alpha | ^2 \ln \frac{| \alpha | ^2}{2}+\frac{| \beta | ^4}{2}  \ln \frac{| \beta | ^4}{2}-| \beta | ^2 \ln \frac{| \beta | ^2}{2}+A_+\ln(A_+)+A_-\ln(A_-)
\end{equation}

\begin{comment}
    
\begin{equation}
\label{mi0akp}
\begin{split}
    %I(\rho_{-k,p}^{(0)})
    I=\Bigg(\frac{1+| \alpha | ^4+| \beta | ^4  -\sqrt{\left(| \alpha | ^4-| \beta | ^4\right)^2+2( | \alpha | ^4+ | \beta | ^4)+1}}{4}\Bigg) \ln \Bigg( \frac{1+| \alpha | ^4+| \beta | ^4  -\sqrt{\left(| \alpha | ^4-| \beta | ^4\right)^2+2( | \alpha | ^4+ | \beta | ^4)+1}}{4}\Bigg)\\+\Bigg(\frac{1+| \alpha | ^4+| \beta | ^4  +\sqrt{\left(| \alpha | ^4-| \beta | ^4\right)^2+2( | \alpha | ^4+ | \beta | ^4)+1}}{4}\Bigg) \ln \Bigg( \frac{1+| \alpha | ^4+| \beta | ^4  +\sqrt{\left(| \alpha | ^4-| \beta | ^4\right)^2+2( | \alpha | ^4+ | \beta | ^4)+1}}{4}\Bigg)\\-(| \alpha | ^2+1) \ln \Bigg(\frac{| \alpha | ^2+1}{2}\Bigg)-| \beta | ^2 \ln \frac{| \beta | ^2}{2}+\Bigg(\frac{| \alpha | ^2 | \beta | ^2}{2}\Bigg) \ln \Bigg( \frac{| \alpha | ^2 | \beta | ^2}{2}\Bigg)
    \end{split}
\end{equation}

\end{comment}

\begin{equation}
\label{mi0akp}
\begin{split}
    %I(\rho_{-k,p}^{(0)})
    I(\rho_{p,-k}^{(0)})=-(| \alpha | ^2+1) \ln \Bigg(\frac{| \alpha | ^2+1}{2}\Bigg)-| \beta | ^2 \ln \frac{| \beta | ^2}{2}+\frac{| \alpha | ^2 | \beta | ^2}{2} \ln \Bigg( \frac{| \alpha | ^2 | \beta | ^2}{2}\Bigg)+B_+\ln{B_+}+B_-\ln{B_-}
    \end{split}
\end{equation}

\begin{comment}

\begin{equation}
\begin{split}
\label{mi0kap}
I=\Bigg(\frac{1+| \alpha | ^4+| \beta | ^4  -\sqrt{\left(| \alpha | ^4-| \beta | ^4\right)^2+2 \left(| \alpha | ^4+| \beta | ^4\right)+1}}{4}\Bigg) \ln \Bigg( \frac{1+| \alpha | ^4+| \beta | ^4  -\sqrt{\left(| \alpha | ^4-| \beta | ^4\right)^2+2 \left(| \alpha | ^4+| \beta | ^4\right)+1}}{4}\Bigg)\\+\Bigg(\frac{1+| \alpha | ^4+| \beta | ^4  +\sqrt{\left(| \alpha | ^4-| \beta | ^4\right)^2+2 \left(| \alpha | ^4+| \beta | ^4\right)+1}}{4}\Bigg) \ln \Bigg(\frac{1+| \alpha | ^4+| \beta | ^4  +\sqrt{\left(| \alpha | ^4-| \beta | ^4\right)^2+2 \left(| \alpha | ^4+| \beta | ^4\right)+1}}{4}\Bigg)\\-| \alpha | ^2 \ln \frac{| \alpha | ^2}{2}-(| \beta | ^2+1) \ln \Bigg(\frac{| \beta | ^2+1}{2}\Bigg)+\Bigg(\frac{| \alpha | ^2 | \beta | ^2}{2} \Bigg)\ln\Bigg(\frac{| \alpha | ^2 | \beta | ^2}{2}\Bigg)
\end{split}  
\end{equation}
\end{comment}

\begin{equation}
\begin{split}
\label{mi0kap}
I(\rho^{(0)}_{-p,k})=-| \alpha | ^2 \ln \frac{| \alpha | ^2}{2}-(| \beta | ^2+1) \ln \Bigg(\frac{| \beta | ^2+1}{2}\Bigg)+\Bigg(\frac{| \alpha | ^2 | \beta | ^2}{2} \Bigg)\ln\Bigg(\frac{| \alpha | ^2 | \beta | ^2}{2}\Bigg)+C_+\ln{C_+}+C_-\ln{C_-}
\end{split}  
\end{equation}

where $$A_\pm=\frac{2 | \alpha | ^2 | \beta | ^2 +1\pm\sqrt{4 | \alpha | ^2 | \beta | ^2+1}}{4},\;B_\pm = \Bigg(\frac{1+| \alpha | ^4+| \beta | ^4  \pm\sqrt{\left(| \alpha | ^4-| \beta | ^4\right)^2+2( | \alpha | ^4+ | \beta | ^4)+1}}{4}\Bigg) $$, \text{and} $$C_\pm=\Bigg(\frac{1+| \alpha | ^4+| \beta | ^4  \pm\sqrt{\left(| \alpha | ^4-| \beta | ^4\right)^2+2 \left(| \alpha | ^4+| \beta | ^4\right)+1}}{4}\Bigg)$$

\begin{figure}[h]
    \centering
    % Subfigure (a)
    \begin{subfigure}[b]{0.32\textwidth}
        \centering
        \includegraphics[scale=0.62]{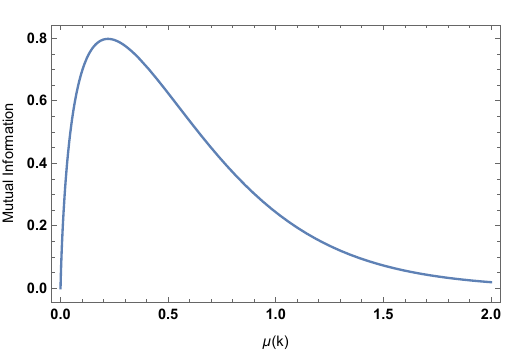}
        \caption{$\rho_{p,k}^{(0)}$ or $\rho_{-p,-k}^{(0)}$}
    \end{subfigure}
    % Subfigure (b)
    \begin{subfigure}[b]{0.32\textwidth}
        \centering
        \includegraphics[scale=0.87]{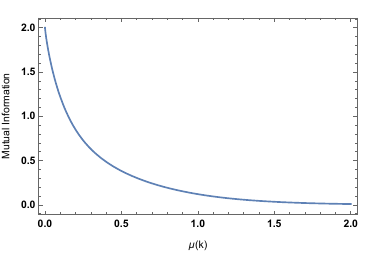}
        \caption{$\rho_{p,-k}^{(0)}$}
    \end{subfigure}
    % Subfigure (c)
    \begin{subfigure}[b]{0.32\textwidth}
        \centering
        \includegraphics[scale=0.87]{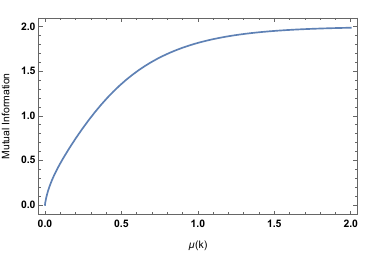}
        \caption{$\rho_{-p,k}^{(0)}$}
    \end{subfigure}
    \caption{\it {\small The mutual information of $\rho_{p,k}^{(0)}$ or $\rho_{-p,-k}^{(0)}$ (a), $\rho_{p,-k}^{(0)}$ (b), and $\rho_{-p,k}^{(0)}$ (c) with respect to the parameter $\mu=(\frac{m^2+2(n_L+1)eB}{eE})$. It is non-monotonic for $\rho_{p,k}^{(0)}$ (or $\rho_{-p,-k}^{(0)}$) and monotonic for $\rho_{p,-k}^{(0)}$ and $\rho_{-p,k}^{(0)}$. It increases with the electric field in the particle-particle (or antiparticle-antiparticle) and particle-antiparticle sectors, while it decreases in the antiparticle-particle sector. In contrast, the magnetic field shows the opposite behavior, influencing the system inversely relative to the electric field. This behavior is consistent with the principle of entanglement monogamy, similar to the violation of Bell's inequality.}}
    \label{fig:MIMF}
\end{figure}

The variation of \ref{mi0kp}(a), \ref{mi0akp}(b) and \ref{mi0kap} (c) with respect to the parameter $\mu$ is shown in \ref{fig:MIMF}. It exhibits non-monotonic behavior in the particle-particle sector, while the particle-antiparticle and antiparticle-particle sectors display monotonic behavior. In the large $\mu$ limit, it vanishes for the particle-particle and particle-antiparticle sectors but attains its maximum value for the antiparticle-particle sector. At this limit, the system reverts to its initial state, as outlined in \ref{Bell-1MF}. Similarly, as observed in the scalar field case, all sectors converge to their respective initial states in the large $\mu$ limit.

\begin{comment}

\begin{figure}[h]
    \centering
\includegraphics[scale=0.61]{figures/MIrho0kp.pdf}
\includegraphics[scale=0.85]{figures/MI0akp.pdf}
\includegraphics[scale=0.85]{figures/MI0kap.pdf}
\hspace{1.0cm}
%\includegraphics[scale=0.62]{figures/LNrho0kp.pdf}
\caption{\it{\small The mutual information of $\rho_{p,k}^{(0)}$ or $\rho_{-p,-k}^{(0)}$ (above left),  $\rho_{p,-k}^{(0)}$ (above right) and $\rho_{-p,k}^{(0)}$ (below left)  vs. $\mu$. It is non-monotonic for $\rho_{p,k}^{(0)}$ and $\rho_{-p,-k}^{(0)}$, and monotonic for $\rho_{p,-k}^{(0)}$ and $\rho_{-p,k}^{(0)}$.}}
    \label{fig:MIMF}
\end{figure}

\end{comment}

\section{Fermions coupled to time-dependent electric field and constant magnetic field}
\label{Fermions coupled to time-dependent electric field and constant magnetic field}
Next, we consider a Sauter-type electric field along the z-direction as
\be E(t)=E_0 \sech^2(t/\tau)\ee
where $\tau$ is the width of the electric field strength. We choose the gauge
\begin{equation}
    \label{Amut}
    A_\mu = \big(0,-By,0,E_0 \tau \tanh{(\frac{t}{\tau})}\big)
\end{equation}

Following the same procedure for the constant field and using the ansatz $\Tilde{\chi}(x)= e^{i(k^x z+k^z z)}\Tilde{\chi}_{s}(x)^{(p)}(t,y)\epsilon_s$, for the particle mode ($p$ stands for particle), we map \ref{eq:EOM1} in this case as
\begin{equation}
    \label{eq:EOM1TD}
    [\partial_t^2-(k^x-eBy)^2+\partial_y^2-(k^z+eE_0 \tau \tanh{(\frac{t}{\tau})})^2-m^2-ie(\gamma^0\gamma^3 E_0\sech^2(\frac{t}{\tau})+\gamma^1\gamma^2B)]\Tilde{\chi}_s(t,y)\epsilon_s=0
\end{equation}
where $\epsilon_s$ are the same as \ref{epsilon_s}. On decoupling \ref{eq:EOM1TD} using $\Tilde{\chi}^{(p)}_{s}(t,y)=h_s(y) \Tilde{f}^{(p)}_s(t)$, the decoupled equations reads as
\begin{equation}
\label{2ndt}
\left(\partial_t^2-(k^z+eE_0 \tau \tanh{(\frac{t}{\tau})})^2-m^2-ieE_0\sech^2(\frac{t}{\tau})\lambda_s-S_s\right)\Tilde{f}^{(p)}_s(t)=0
\end{equation}
\begin{equation}
    \label{2ndy}
\left(\partial_y^2-(k^x-eBy)^2-ieB\beta_s+S_s\right)h_s(y)=-0
\end{equation}

Here, $S_s$ is the separation coefficient, the same as the constant electric field case, and eigenvalues $\lambda_s$ and $\beta_s$ are the same as the constant electric field case.

Following \cite{Ebadi:2014ufa} and \cite{AS}, the solution of second order differential equation of time \ref{2ndt} is given by
\begin{equation}
    \Tilde{f}^{(p)}_s(t)=(\xi-1)^{i\tau \omega_{k,\text{in}/2}}\xi^{-i\tau \omega_{k,\text{out}}/2} [C_1 F(a,b,c;\xi)+C_2\xi^{1-c}F(a-c+1,b-c+1,2-c;\xi)]
\end{equation}
where
$$\omega_{k,\text{in}}=\sqrt{(k^z-eE_0\tau)^2+m^2+S_s},\quad \omega_{k,\text{out}}=\sqrt{(k^z+eE_0\tau)^2+m^2+S_s},$$ 
$$\xi=\frac{1}{2}\tanh{\frac{t}{\tau}}+\frac{1}{2},\quad
     \rho=eE_0\tau^2,\quad
     a=\frac{1}{2}\Big(1+i(\tau \omega_{k,\text{in}}-\tau \omega_{k,\text{out}})\Big)\pm i \rho,$$

$$ b=1+\frac{i}{2}\Big(\tau \omega_{k,\text{in}}-\tau \omega_{k,\text{out}}\Big)\mp i \rho,\;\text{and}\;
     c=1-i\tau \omega_{k,\text{out}} $$

Next, the solution of \ref{2ndy} is the same as \ref{YTI}. Similar to the constant electric field case full fermionic modes are given as
\begin{equation}
\begin{split}
\label{modef1TD}
     U^{n_L, \text{in}}_{k,s}(x)  =\frac{1}{R_s}\left(i \gamma^{\mu} \partial_{\mu}-e \gamma^{\mu} A_{\mu}+m\right) e^{-i(k^x x+k^z z)}e^{-{y}_+^2/2}
	H_{n_L}({y}_-)
	(\xi-1)^{i\tau \omega_{k,\text{in}/2}}\xi^{-i\tau \omega_{k,\text{out}}/2}F(a,b,c;\xi) \epsilon_{s}\\ \quad (s=1,2)
 \end{split}
\end{equation}
\begin{equation}
\begin{split}
\label{modef2TD}
V^{n_L, \text{in}}_{k,s}(x)  =\frac{1}{R_s}\left(i \gamma^{\mu} \partial_{\mu}-e \gamma^{\mu} A_{\mu}+m\right) e^{i(k^x x+k^z z)}e^{-{y}_-^2/2}
	H_{n_L}({y}_+)[(\xi-1)^{i\tau \omega_{k,\text{in}/2}}\xi^{-i\tau \omega_{k,\text{out}}/2} F(a,b,c;\xi)]^* \epsilon_{s} \\ \quad (s=3,4)    
 \end{split}
\end{equation}
\begin{equation}
\begin{split}
    \label{modef3TD}
 U^{n_L, \text{out}}_{k,s}(x)  =\frac{1}{R_s}\left(i \gamma^{\mu} \partial_{\mu}-e \gamma^{\mu} A_{\mu}+m\right) e^{-i(k^x x+k^z z)}e^{-{y}_+^2/2}
	H_{n_L}({y}_-) (\xi-1)^{i\tau \omega_{k,\text{in}/2}}\xi^{-i\tau \omega_{k,\text{out}}/2} \xi^{1-c}\\ \times F(a-c+1,b-c+1,2-c;\xi)\epsilon_{s} \qquad (s=1,2)
 \end{split}
\end{equation}
\begin{equation}
\begin{split}
\label{modef4TD}
   V^{n_L, \text{out}}_{k,s}(x)  =\frac{1}{R_s}\left(i \gamma^{\mu} \partial_{\mu}-e \gamma^{\mu} A_{\mu}+m\right) e^{i(k^x x+k^z z)}e^{-{y}_-^2/2}
	H_{n_L}({y}_+)
	[(\xi-1)^{i\tau \omega_{k,\text{in}/2}}\xi^{-i\tau \omega_{k,\text{out}}/2}\xi^{1-c}\\ \times F(a-c+1,b-c+1,2-c;\xi)]^* \epsilon_{s} \quad \quad (s=3,4) 
 \end{split}
\end{equation}

Here, $U_{k,s}^{n_L}(x)$ and $V_{k,s}^{n_L}(x)$ represent modes for particles and antiparticles, respectively, with $R_s$ serving as the normalization factor. The orthogonality relations between these modes, the canonical field quantization and the Bogoliubov transformation remain consistent with those observed in the constant electric field case.

\begin{figure}[h]
    \centering
    % Subfigure (a)
    \begin{subfigure}[b]{0.32\textwidth}
        \centering
        \includegraphics[scale=0.50]{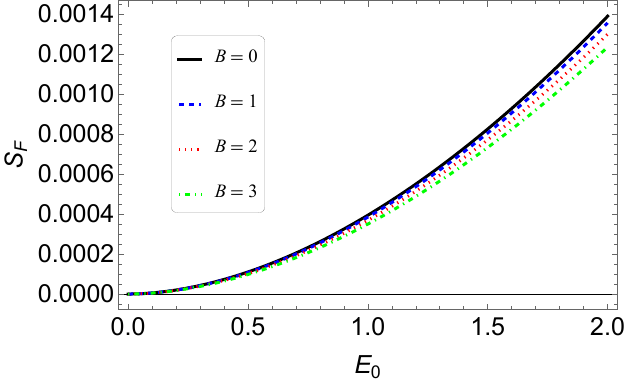}
        \caption{\small  \( S_F \) vs $E_0$, where, \(e = 0.1\) and \(n_L, B, k = 1\).}
    \end{subfigure}\hspace{3cm}
    % Subfigure (b)
    \begin{subfigure}[b]{0.32\textwidth}
        \centering
        \includegraphics[scale=0.70]{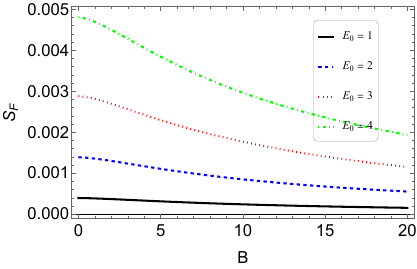}
        \caption{\small   \( S_F \) vs $B$, where, \(e = 0.1\) and \(n_L, E_0, k = 1\).}
    \end{subfigure}
    \\
    % Subfigure (c)
    \begin{subfigure}[b]{0.32\textwidth}
        \centering
        \includegraphics[scale=0.70]{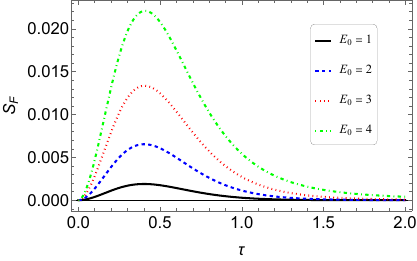}
        \caption{\small  \( S_F \) vs $\tau$, where, \(e = 0.1\) and \(n_L, B, m, k = 1\).}
    \end{subfigure}\hspace{3cm}
    % Subfigure (d)
    \begin{subfigure}[b]{0.32\textwidth}
        \centering
        \includegraphics[scale=0.70]{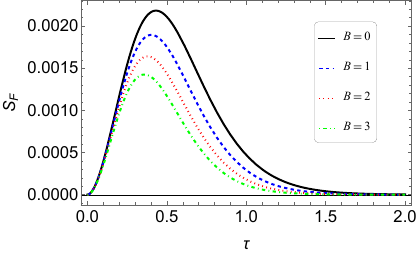}
        \caption{\small   \( S_F\) vs $\tau$, where, \(e = 0.1\) and \(n_L, E_0, m, k = 1\).}
    \end{subfigure}
    \\
    % Subfigure (e)
    \begin{subfigure}[b]{0.32\textwidth}
        \centering
        \includegraphics[scale=0.70]{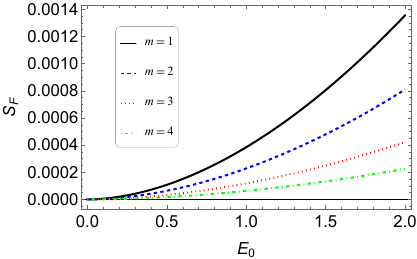}
        \caption{\small   \( S_F \) vs $E_0$, where, \(e, \tau = 0.1\) and \(n_L, k, B = 1\).}
    \end{subfigure}\hspace{3cm}
    % Subfigure (f)
    \begin{subfigure}[b]{0.32\textwidth}
        \centering
        \includegraphics[scale=0.70]{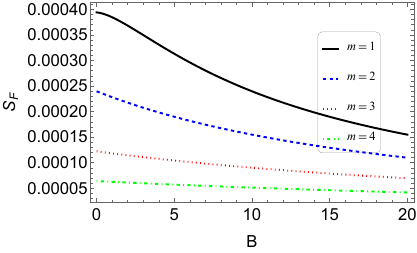}
        \caption{\small   \( S_F \) vs $B$, where, \(e, \tau = 0.1\) and \(n_L, k, E_0 = 1\).}
    \end{subfigure}
    \\
    % Subfigure (g)
    \begin{subfigure}[b]{0.32\textwidth}
        \centering
        \includegraphics[scale=0.70]{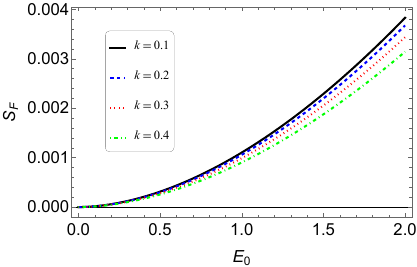}
        \caption{\small   \( S_F \) vs $E_0$, where, \(e, \tau = 0.1\) and \(n_L, B, m = 1\).}
    \end{subfigure}\hspace{3cm}
    % Subfigure (h)
    \begin{subfigure}[b]{0.32\textwidth}
        \centering
        \includegraphics[scale=0.70]{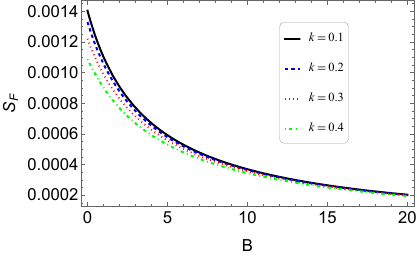}
        \caption{\small   \( S_F \) vs $B$, where, \(e, \tau = 0.1\) and \(n_L, E_0, m = 1\).}
    \end{subfigure}
    \caption{\it {\small Under the influence of a time-dependent electric field, the entanglement entropy ($S_F$) between particles and antiparticles in the vacuum is analyzed as a function of various parameters while keeping others fixed. For a fixed electric and magnetic field, the variation of $S_F$ with respect to $\tau$ exhibits local maxima. In contrast, all other plots show a monotonic behavior, with $S_F$ increasing as $E_0$ increases and decreasing with an increase in the magnetic field. Additionally, it is observed that $S_F$
  decreases as the mass $m$ and momentum $k$ increase. The parameters used in the accompanying figure are $e=0.1$, $E_0=1$, $\tau=0.1$, $n_L=1$, $m=1$, $k=1$, and $B=1$. Similarly to \ref{fig:FNDM}, the violation of Bell's inequality follows a trend comparable to that of the entanglement entropy, with values exceeding $2$.}}
    \label{SFTD}
\end{figure}

\begin{figure}[h]
    \centering
    % Subfigure (a)
    \begin{subfigure}[b]{0.32\textwidth}
        \centering
        \includegraphics[scale=0.70]{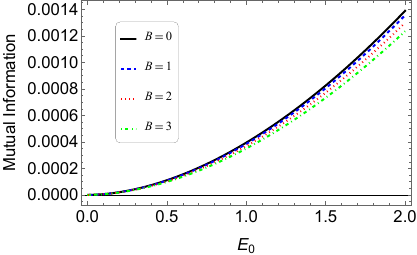}
        \caption{\small  \( I(\rho^{(0)}_{p,k}) \) vs $E_0$, where, \(e, \tau = 0.1\) and \(n_L, m, k = 1\).}
    \end{subfigure}\hspace{3cm}
    % Subfigure (b)
    \begin{subfigure}[b]{0.32\textwidth}
        \centering
        \includegraphics[scale=0.70]{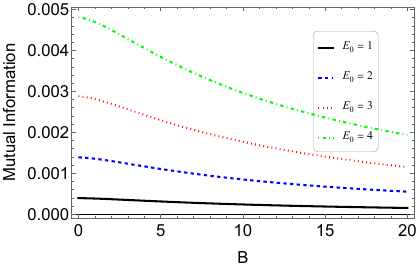}
        \caption{\small   \( I(\rho^{(0)}_{p,k}) \) vs $B$, where, \(e, \tau = 0.1\) and \(n_L, m, k = 1\).}
    \end{subfigure}
    \\
    % Subfigure (c)
    \begin{subfigure}[b]{0.32\textwidth}
        \centering
        \includegraphics[scale=0.70]{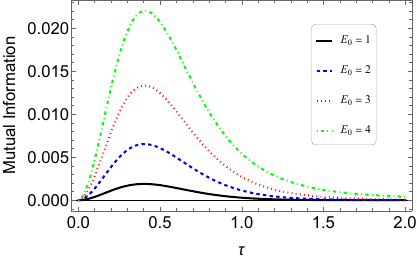}
        \caption{\small  \( I(\rho^{(0)}_{p,k}) \) vs $\tau$, where, \(e = 0.1\) and \(n_L, m, k, B = 1\).}
    \end{subfigure}\hspace{3cm}
    % Subfigure (d)
    \begin{subfigure}[b]{0.32\textwidth}
        \centering
        \includegraphics[scale=0.75]{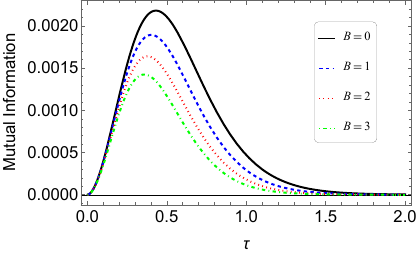}
        \caption{\small   \( I(\rho^{(0)}_{p,k}) \) vs $\tau$, where, \(e = 0.1\) and \(n_L, m, k,E_0 = 1\).}
    \end{subfigure}
    \\
    % Subfigure (e)
    \begin{subfigure}[b]{0.32\textwidth}
        \centering
        \includegraphics[scale=0.70]{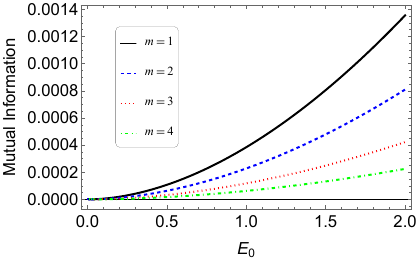}
        \caption{\small   \( I(\rho^{(0)}_{p,k}) \) vs $E_0$, where, \(e, \tau = 0.1\) and \(n_L, B, k = 1\).}
    \end{subfigure}\hspace{3cm}
    % Subfigure (f)
    \begin{subfigure}[b]{0.32\textwidth}
        \centering
        \includegraphics[scale=0.70]{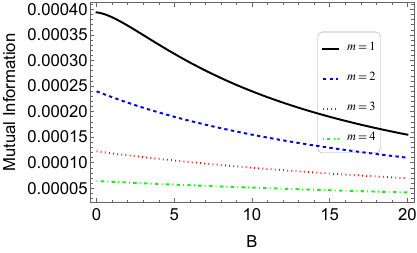}
        \caption{\small   \( I(\rho^{(0)}_{p,k}) \) vs $B$, where, \(e, \tau = 0.1\) and \(n_L, E_0, k = 1\).}
    \end{subfigure}
    \\
    % Subfigure (g)
    \begin{subfigure}[b]{0.32\textwidth}
        \centering
        \includegraphics[scale=0.70]{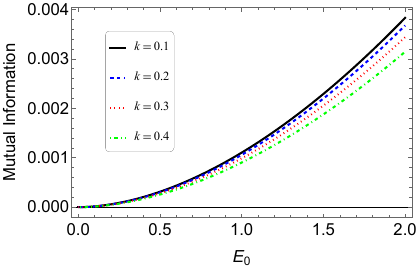}
        \caption{\small   \( I(\rho^{(0)}_{p,k}) \) vs $E_0$, where, \(e, \tau = 0.1\) and \(n_L, m, B = 1\).}
    \end{subfigure}\hspace{3cm}
    % Subfigure (h)
    \begin{subfigure}[b]{0.32\textwidth}
        \centering
        \includegraphics[scale=0.70]{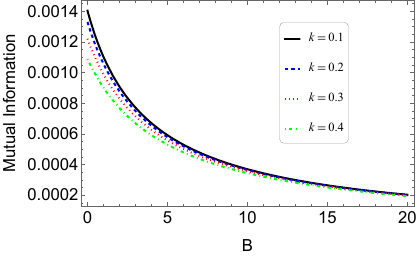}
        \caption{\small   \( I(\rho^{(0)}_{p,k}) \) vs $B$, where, \(e, \tau = 0.1\) and \(n_L, m, E_0 = 1\).}
    \end{subfigure}
     \caption{\it {\small We analyze the mutual information \( I(\rho^{(0)}_{p,k}) \) for the particle-particle sector in the zero charge state under the influence of a time-dependent electric field while keeping other parameters constant. For fixed electric and magnetic fields, \( I(\rho^{(0)}_{p,k}) \) varies with \(\tau\), displaying local maxima. In contrast, other plots exhibit monotonic behavior, where \( I(\rho^{(0)}_{p,k}) \) increases with growing \( E_0 \) and decreases as the magnetic field strength increases. Furthermore, \( I(\rho^{(0)}_{p,k}) \) is observed to decrease with increasing mass \( m \) and momentum \( k \). Parameters used in the accompanying figure are \( e = 0.1 \), \( E_0 = 1 \), \( \tau = 0.1 \), \( n_L = 1 \), \( m = 1 \), \( k = 1 \), and \( B = 1 \).
}}
    \label{I0PKTD}
\end{figure}

\begin{figure}[h]
    \centering
    % Subfigure (a)
    \begin{subfigure}[b]{0.32\textwidth}
        \centering
        \includegraphics[scale=0.70]{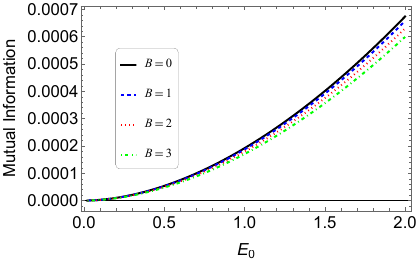}
        \caption{\small  \( I(\rho^{(0)}_{p,-k}) \) vs $E_0$, where, \(e, \tau = 0.1\) and \(n_L, m, k = 1\).}
    \end{subfigure}\hspace{3cm}
    % Subfigure (b)
    \begin{subfigure}[b]{0.32\textwidth}
        \centering
        \includegraphics[scale=0.70]{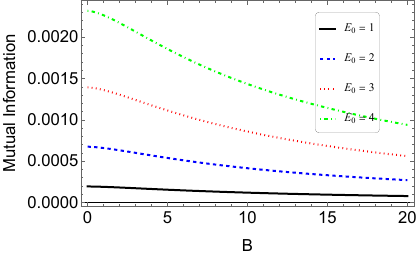}
        \caption{\small   \( I(\rho^{(0)}_{p,-k}) \) vs $B$, where, \(e, \tau = 0.1\) and \(n_L, m, k = 1\).}
    \end{subfigure}
    \\
    % Subfigure (c)
    \begin{subfigure}[b]{0.32\textwidth}
        \centering
        \includegraphics[scale=0.70]{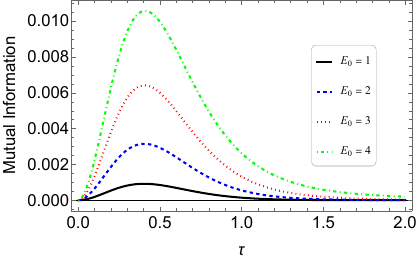}
        \caption{\small  \( I(\rho^{(0)}_{p,-k}) \) vs $\tau$, where, \(e = 0.1\) and \(n_L, m, k, B = 1\).}
    \end{subfigure}\hspace{3cm}
    % Subfigure (d)
    \begin{subfigure}[b]{0.32\textwidth}
        \centering
        \includegraphics[scale=0.75]{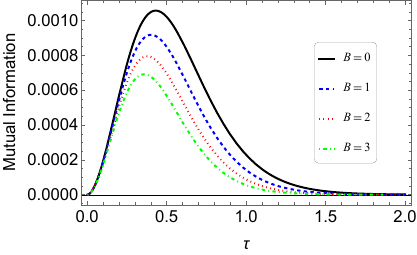}
        \caption{\small   \( I(\rho^{(0)}_{p,-k}) \) vs $\tau$, where, \(e = 0.1\) and \(n_L, m, k, E_0 = 1\).}
    \end{subfigure}
    \\
    % Subfigure (e)
    \begin{subfigure}[b]{0.32\textwidth}
        \centering
        \includegraphics[scale=0.70]{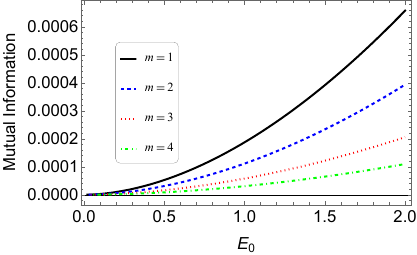}
        \caption{\small   \( I(\rho^{(0)}_{p,-k}) \) vs $E_0$, where, \(e, \tau = 0.1\) and \(n_L, B, k = 1\).}
    \end{subfigure}\hspace{3cm}
    % Subfigure (f)
    \begin{subfigure}[b]{0.32\textwidth}
        \centering
        \includegraphics[scale=0.70]{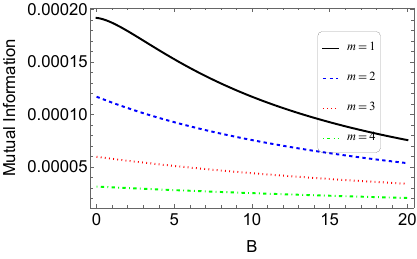}
        \caption{\small   \( I(\rho^{(0)}_{p,-k}) \) vs $B$, where, \(e, \tau = 0.1\) and \(n_L, E_0, k = 1\).}
    \end{subfigure}
    \\
    % Subfigure (g)
    \begin{subfigure}[b]{0.32\textwidth}
        \centering
        \includegraphics[scale=0.70]{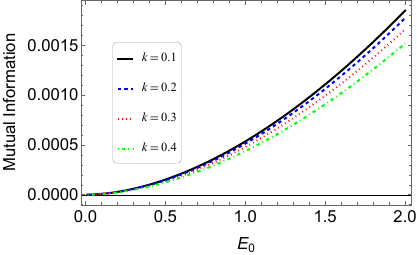}
        \caption{\small   \( I(\rho^{(0)}_{p,-k}) \) vs $E_0$, where, \(e, \tau = 0.1\) and \(n_L, m, B = 1\).}
    \end{subfigure}\hspace{3cm}
    % Subfigure (h)
    \begin{subfigure}[b]{0.32\textwidth}
        \centering
        \includegraphics[scale=0.70]{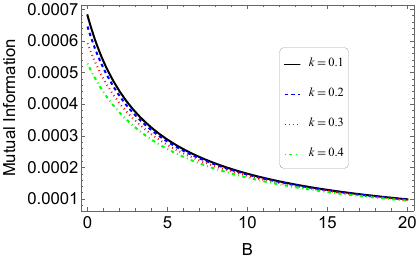}
        \caption{\small   \( I(\rho^{(0)}_{p,-k}) \) vs $B$, where, \(e, \tau = 0.1\) and \(n_L, m, E_0 = 1\).}
    \end{subfigure}
    \caption{\it {\small We analyze the mutual information \( I(\rho^{(0)}_{p,-k}) \) for the particle-antiparticle sector in the zero charge state under the influence of a time-dependent electric field while keeping other parameters fixed. For a given electric and magnetic field, \( I(\rho^{(0)}_{p,-k}) \) varies with \(\tau\), showing local maxima. In contrast, other plots exhibit monotonic behavior, with \( I(\rho^{(0)}_{p,-k}) \) increasing as \( E_0 \) grows and decreasing with higher magnetic field values. Moreover, \( I(\rho^{(0)}_{p,-k}) \) decreases as the mass \( m \) and momentum \( k \) increase. Parameters used in the accompanying figure are \( e = 0.1 \), \( E_0 = 1 \), \( \tau = 0.1 \), \( n_L = 1 \), \( m = 1 \), \( k = 1 \), and \( B = 1 \). The variation of \( I(\rho^{(0)}_{p,-k}) \) closely resembles that of \( I(\rho^{(0)}_{p,k}) \).
}}
    \label{I0APKTD}
\end{figure}

\begin{figure}[h]
    \centering
    % Subfigure (a)
    \begin{subfigure}[b]{0.32\textwidth}
        \centering
        \includegraphics[scale=0.70]{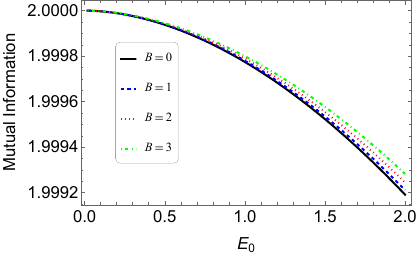}
        \caption{\small  \( I(\rho^{(0)}_{-p,k}) \) vs $E_0$, where, \(e, \tau = 0.1\) and \(n_L, m, k = 1\).}
    \end{subfigure}\hspace{3cm}
    % Subfigure (b)
    \begin{subfigure}[b]{0.32\textwidth}
        \centering
        \includegraphics[scale=0.70]{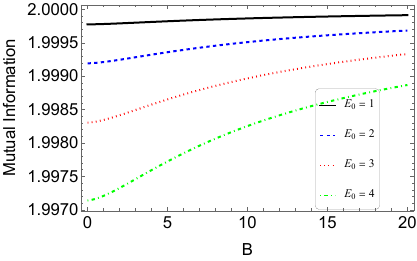}
        \caption{\small   \( I(\rho^{(0)}_{-p,k}) \) vs $B$, where, \(e, \tau = 0.1\) and \(n_L, m, k = 1\).}
    \end{subfigure}
    \\
    % Subfigure (c)
    \begin{subfigure}[b]{0.32\textwidth}
        \centering
        \includegraphics[scale=0.70]{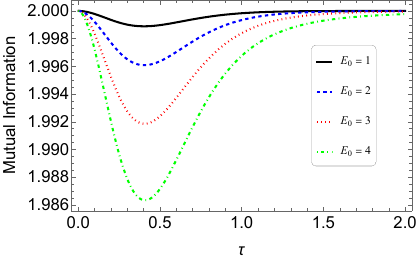}
        \caption{\small  \( I(\rho^{(0)}_{-p,k}) \) vs $\tau$, where, \(e = 0.1\) and \(n_L, m, k, B = 1\).}
    \end{subfigure}\hspace{3cm}
    % Subfigure (d)
    \begin{subfigure}[b]{0.32\textwidth}
        \centering
        \includegraphics[scale=0.75]{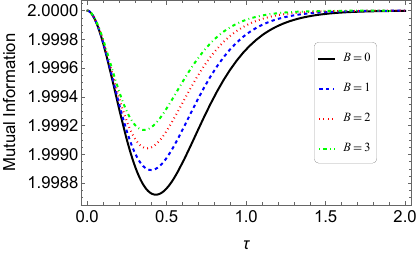}
        \caption{\small   \( I(\rho^{(0)}_{-p,k}) \) vs $\tau$, where, \(e = 0.1\) and \(n_L, m, k, E_0 = 1\).}
    \end{subfigure}
    \\
    % Subfigure (e)
    \begin{subfigure}[b]{0.32\textwidth}
        \centering
        \includegraphics[scale=0.70]{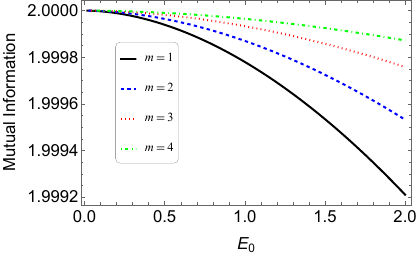}
        \caption{\small   \( I(\rho^{(0)}_{-p,k}) \) vs $E_0$, where, \(e, \tau = 0.1\) and \(n_L, B, k = 1\).}
    \end{subfigure}\hspace{3cm}
    % Subfigure (f)
    \begin{subfigure}[b]{0.32\textwidth}
        \centering
        \includegraphics[scale=0.70]{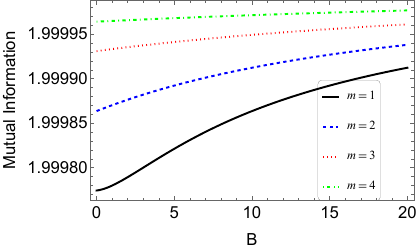}
        \caption{\small   \( I(\rho^{(0)}_{-p,k}) \) vs $B$, where, \(e, \tau = 0.1\) and \(n_L, E_0, k = 1\).}
    \end{subfigure}
    \\
    % Subfigure (g)
    \begin{subfigure}[b]{0.32\textwidth}
        \centering
        \includegraphics[scale=0.70]{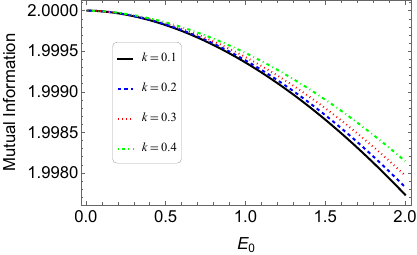}
        \caption{\small   \( I(\rho^{(0)}_{-p,k}) \) vs $E_0$, where, \(e, \tau = 0.1\) and \(n_L, m, B = 1\).}
    \end{subfigure}\hspace{3cm}
    % Subfigure (h)
    \begin{subfigure}[b]{0.32\textwidth}
        \centering
        \includegraphics[scale=0.70]{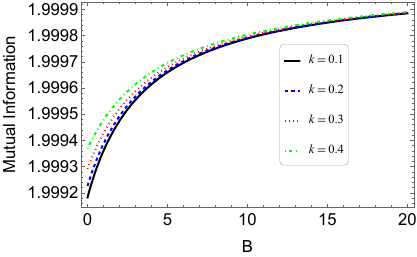}
        \caption{\small   \( I(\rho^{(0)}_{-p,k}) \) vs $B$, where, \(e, \tau = 0.1\) and \(n_L, m, E_0 = 1\).}
    \end{subfigure}
    \caption{\it {\small We analyze the mutual information \( I(\rho^{(0)}_{-p,k}) \) in the antiparticle-particle sector for the zero charge state under the influence of a time-dependent electric field, while keeping other parameters constant. For a fixed electric and magnetic field, the variation of \( I(\rho^{(0)}_{-p,k}) \) with respect to \(\tau\) exhibits local maxima at small \(\tau\) values. In contrast, all other plots display monotonic behavior, with \( I(\rho^{(0)}_{-p,k}) \) decreasing as \(E_0\) increases and increasing with higher magnetic field values. Furthermore, \( I(\rho^{(0)}_{-p,k}) \) is observed to grow with increasing mass \(m\) and momentum \(k\). The parameters used in the accompanying figure are \(e = 0.1\), \(E_0 = 1\), \(\tau = 0.1\), \(n_L = 1\), \(m = 1\), \(k = 1\), and \(B = 1\).
}}
    \label{I0PAKTD}
\end{figure}

\begin{figure}[h]
    \centering
    % Subfigure (a)
    \begin{subfigure}[b]{0.32\textwidth}
        \centering
        \includegraphics[scale=0.50]{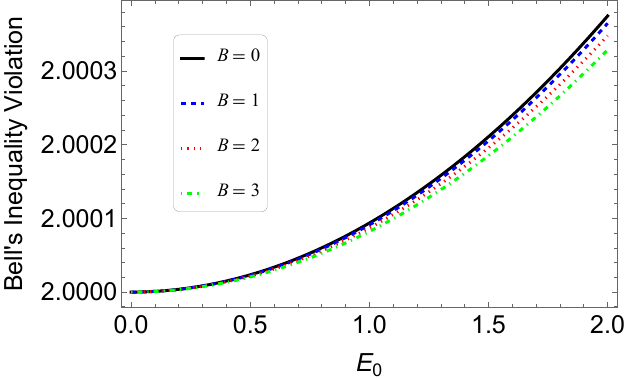}
        \caption{\small \(\langle \mathcal{B}_{\rm max}\rangle\) vs $E_0$, where, \(e, \tau = 0.1\) and \(n_L, m, k = 1\).}
    \end{subfigure}\hspace{3cm}
    % Subfigure (b)
    \begin{subfigure}[b]{0.32\textwidth}
        \centering
        \includegraphics[scale=0.50]{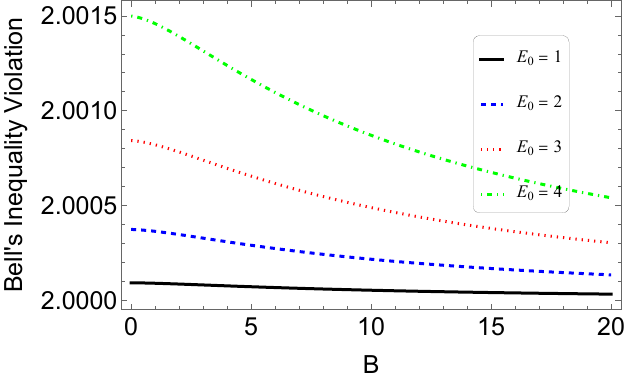}
        \caption{\small  \(\langle \mathcal{B}_{\rm max}\rangle\) vs $B$, where, \(e, \tau = 0.1\) and \(n_L, m, k = 1\).}
    \end{subfigure}
    \\
    % Subfigure (c)
    \begin{subfigure}[b]{0.32\textwidth}
        \centering
        \includegraphics[scale=0.50]{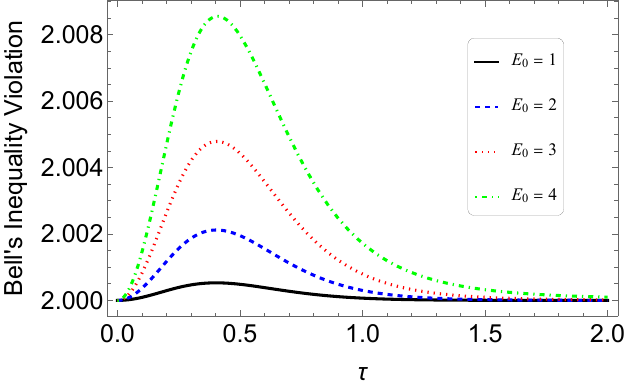}
        \caption{\small  \(\langle \mathcal{B}_{\rm max}\rangle\) vs $\tau$, where, \(e = 0.1\) and \(n_L, m, k, B = 1\).}
    \end{subfigure}\hspace{3cm}
    % Subfigure (d)
    \begin{subfigure}[b]{0.32\textwidth}
        \centering
        \includegraphics[scale=0.50]{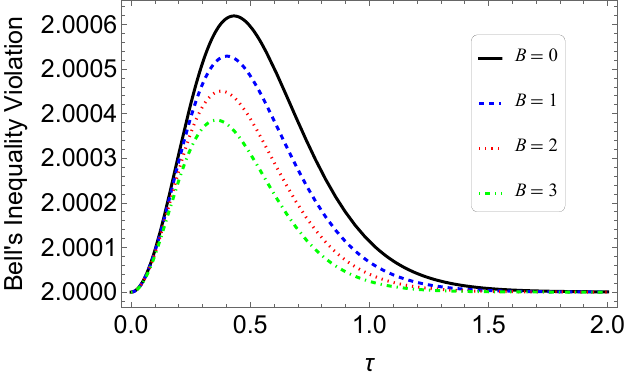}
        \caption{\small  \(\langle \mathcal{B}_{\rm max}\rangle\) vs $\tau$, where, \(e = 0.1\) and \(n_L, m, k, E_0 = 1\).}
    \end{subfigure}
    \\
    % Subfigure (e)
    \begin{subfigure}[b]{0.32\textwidth}
        \centering
        \includegraphics[scale=0.50]{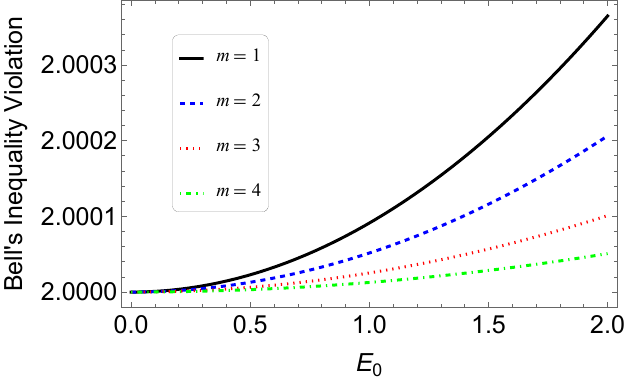}
        \caption{\small  \(\langle \mathcal{B}_{\rm max}\rangle\) vs $E_0$, where, \(e, \tau = 0.1\) and \(n_L, k, B = 1\).}
    \end{subfigure}\hspace{3cm}
    % Subfigure (f)
    \begin{subfigure}[b]{0.32\textwidth}
        \centering
        \includegraphics[scale=0.50]{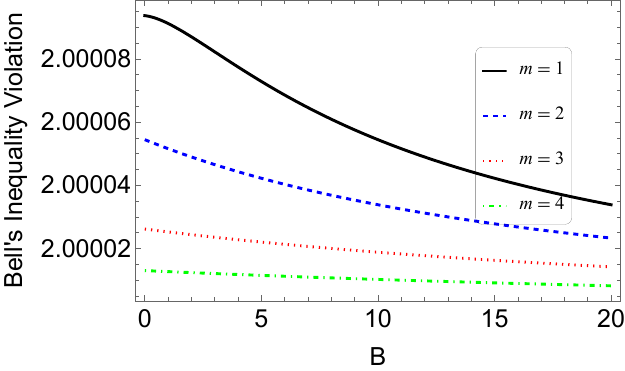}
        \caption{\small  \(\langle \mathcal{B}_{\rm max}\rangle\) vs $B$, where, \(e, \tau = 0.1\) and \(n_L, k, E_0 = 1\).}
    \end{subfigure}
    \\
    % Subfigure (g)
    \begin{subfigure}[b]{0.32\textwidth}
        \centering
        \includegraphics[scale=0.50]{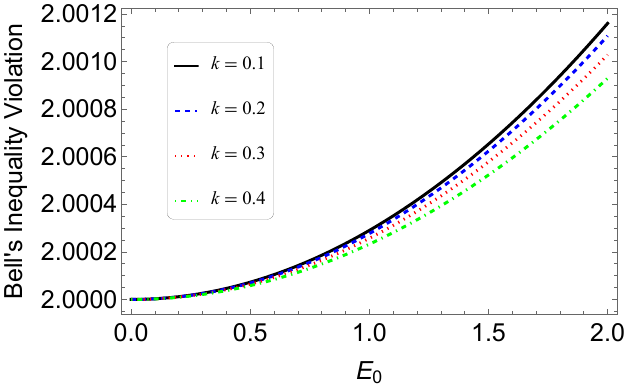}
        \caption{\small  \(\langle \mathcal{B}_{\rm max}\rangle\) vs $E_0$, where, \(e, \tau = 0.1\) and \(n_L, B, m = 1\).}
    \end{subfigure}\hspace{3cm}
    % Subfigure (h)
    \begin{subfigure}[b]{0.32\textwidth}
        \centering
        \includegraphics[scale=0.50]{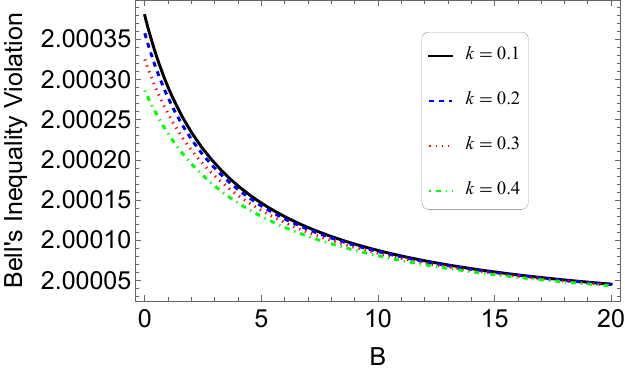}
        \caption{\small  \(\langle \mathcal{B}_{\rm max}\rangle\) vs $B$, where, \(e, \tau = 0.1\) and \(n_L, E_0, m = 1\).}
    \end{subfigure}
    \caption{\it {\small We analyze the violation of Bell's inequality \((\rho^{(0)}_{-p,k})\) for antiparticle-particle pairs in the zero charge state under the influence of a time-dependent electric field, while keeping other parameters constant. For a fixed electric and magnetic field, its variation with respect to \(\tau\) exhibits local maxima at small \(\tau\) values. In contrast, all other plots display monotonic behavior, with \(\langle \mathcal{B}_{\rm max}\rangle (\rho^{(0)}_{-p,k})\) increasing as \(E_0\) increases and decreasing as \(B\) increases. Furthermore, \(\langle \mathcal{B}_{\rm max}\rangle (\rho^{(0)}_{-p,k})\) is observed to decay with increasing mass \(m\) and momentum \(k\). The parameters used in the accompanying figure are \(e = 0.1\), \(E_0 = 1\), \(\tau = 0.1\), \(n_L = 1\), \(m = 1\), \(k = 1\), and \(B = 1\).}}
    \label{B0APKTD}
\end{figure}

Following \cite{Ebadi:2014ufa} and using the orthogonality conditions and the properties of confluent hypergeometric functions \cite{AS} one can find Bogoliubov coefficients as follows

\begin{equation}
    \label{BCTD}
    \begin{split}
        |\beta_{k,s}|^2=\frac{\cosh{(2\pi\rho)}-\cosh{\pi \tau (\omega_{\text{in}}-\omega_{\text{out}})} } {2\sinh{(\pi \tau \omega_{\text{out}})}\sinh{(\pi \tau \omega_{\text{in}})}}\\
         |\alpha_{k,s}|^2=\frac{\cosh{\pi \tau (\omega_{\text{in}}+\omega_{\text{out}})} -\cosh{(2\pi\rho)}} {2\sinh{(\pi \tau \omega_{\text{out}})}\sinh{(\pi \tau \omega_{\text{in}})}}
    \end{split}
\end{equation}
The same as in the constant electric field case, the entanglement between particles and antiparticles is given by \ref{entropyMF}. Here also, due to the spin multiplicity, we consider only a single spin. The variation of the entanglement entropy with respect to the electric field and magnetic field strength is shown in \ref{SFTD}, \ref{B0APKTD}, \ref{I0PKTD}, \ref{I0APKTD} and \ref{I0PAKTD}. However, unlike the previous case where a parameter like $\mu$ is fixed, in this case, we analyze the variation of the entanglement measure with respect to different parameters while keeping all others constant.

\section{Summary and outlook}
\label{Summary and outlook}

We now aim to summarize our findings. The primary objective of this work was to assess the impact of a background magnetic field on the quantum correlations between fermionic Schwinger pairs generated by both constant and Sauter-type electric fields. We examined the vacuum number density, entanglement entropy, and Bell's inequality violation in \ref{ND}. In \ref{Maximally entangled state}, we calculated the quantum mutual information and Bell's inequality violation for a maximally entangled state with zero electric charge. We highlighted the qualitative differences in the behaviour of the information quantities across different sectors of these states.
It is also worth noting that since the number density of created particles is given by  $|\beta_k|^2=e^{-\pi \mu}$, all the plots above will exhibit similar behaviour concerning $|\beta_k|^2$. Following this, we extended these results to a Sauter-type electric field in  \ref{Fermions coupled to time-dependent electric field and constant magnetic field}.

Similarly to the scalar field case discussed in \cite{HSSS}, we observe that in all the plots, the various information quantities converge to specific points for sufficiently large $\mu(k)$ values. Assuming, for instance, that the mass, electric field, and Landau level are fixed, a large $\mu(k)$ corresponds to large values of the magnetic field. In this limit, the Bogoliubov transformation simplifies, resulting in the 'out' state coinciding with the 'in' state, apart from a trivial phase factor. However, unlike the case of the scalar field, we still observe non-zero correlations for the vacuum state, even in this limit. Additionally, the quantum correlation for the vacuum state of the fermionic field is very different, in contrast to those of the scalar field. At the optimal value of $\mu(k)$ observed for which the entanglement entropy and Bell's inequality violation would be maximized \ref{fig:FNDM}. For different sectors of a maximally entangled state, a violation of Bell's inequality is observed only in the antiparticle-particle sector \ref{FigB0apk}, while all other sectors do not show any violation. The variation of mutual information across different sectors is shown in \ref{fig:MIMF}. We find that the mutual information is non-monotonic for the particle-particle and antiparticle-antiparticle sectors, while it is monotonic for the particle-antiparticle and antiparticle-particle sectors.

For a time-dependent electric field, the behaviour of entanglement generated by fermionic modes with respect to $E_0$, $B$, $m$, and $k_z$ is similar to what is observed in the case of a constant electric field. However, for small values of $\tau$ and different $E_0$ values, the entanglement shows a local maximum \ref{SFTD}. High-power laser pulses are promising for experimentally generating entangled states. Furthermore, for the maximally entangled state, we observe behaviour similar to that of a constant electric field.

In the future, we plan to explore quantum correlations through other forms of electric and magnetic fields. An interesting avenue for further research is investigating how entanglement is generated by background electromagnetic fields at finite temperatures.


%%%%%%%%%%%%%%%%%%%%%%%%%%%%%%%%%%%%%%

%%%%%%%%%%%%%%%%%%%%%%%%%%%%%%%%%%%%%


\begin{thebibliography}{99}

\bibitem{bell_1}
A.~Einstein, B.~Podolsky and N.~Rosen, {\it Can Quantum-Mechanical Description of Physical Reality Be Considered Complete}, Phys. Rev. \textbf{777} (1935) 

\bibitem{bell_2}
S.~Bell, {\it On the Einstein-Podolsky-Rosen paradox}, Physics\textbf{1}  195 (1964)

\bibitem{bell_3}
J.~F.~Clauser, M.~A.~Horne, A.~Shimony and R.~A.~Holt, {\it Proposed experiment to test local hidden-variable theories}, Phys.~Rev.~Lett.\textbf{23} 880 (1969) 

\bibitem{bell_4} 
  R.~F.~Werner,
  {\it Quantum states with Einstein-Podolsky-Rosen correlations admitting a hidden-variable model},
   Phys.\ Rev.\ A {\bf 40}, 4277 (1989)

   \bibitem{Aspect1}
A.~Aspect, P.~Grangier and G.~Roger, {\it Experimental Tests of Realistic Local Theories via Bell's Theorem},
Phys.~Rev.~Lett.{\bf47}, 460 (1981)

\bibitem{Aspect2}
A.~Aspect, J.~Dalibard and G.~Roger, {\it Experimental test of Bell's inequalities using time varying analyzers},
Phys. Rev. Lett.49, 1804 (1982)


\bibitem{Mukohyama:1996yi}
S.~Mukohyama, M.~Seriu and H.~Kodama,
{\it Can the entanglement entropy be the origin of black hole entropy?},
Phys. Rev. D \textbf{55}, 7666-7679 (1997)
%doi:10.1103/PhysRevD.55.7666
[arXiv:gr-qc/9701059 [gr-qc]]

\bibitem{MartinMartinez:2010ar}
E.~Martin-Martinez, L.~J.~Garay and J.~Leon,
{\it Unveiling quantum entanglement degradation near a Schwarzschild black hole},
Phys. Rev. D \textbf{82}, 064006 (2010)
[arXiv:1006.1394 [quant-ph]]

\bibitem{Horowitz:2003he}
G.~T.~Horowitz and J.~M.~Maldacena,
{\it The Black hole final state},
JHEP \textbf{02}, 008 (2004)
[arXiv:hep-th/0310281 [hep-th]]

\bibitem{Alsing:2006cj}
P.~M.~Alsing, I.~Fuentes-Schuller, R.~B.~Mann and T.~E.~Tessier,
{\it Entanglement of Dirac fields in non-inertial frames},
Phys. Rev. A \textbf{74}, 032326 (2006)
%doi:10.1103/PhysRevA.74.032326
[arXiv:quant-ph/0603269 [quant-ph]]

\bibitem{Adesso:2007gm}
G.~Adesso and I.~Fuentes-Schuller,
{\it Correlation loss and multipartite entanglement across a black hole horizon},
Quant. Inf. Comput. \textbf{9}, no.7-8, 0657-0665 (2009)
%doi:10.26421/QIC9.7-8-8
[arXiv:quant-ph/0702001 [quant-ph]]



%%%%%%%%%%%%%%%%%%%%%%%%%%%%%%%%%%%%%%%%%%%%%%%%%%%%%%%%%%%%%%%%%%vacuum fluctuations%%%%%%%%%%%%%%%%%%%%%%%%%%%%%%%%%%%%%%%%%%%%%%% 

\bibitem{Reynaud:2001kc}
S.~Reynaud, A.~Lambrecht, C.~Genet and M.~T.~Jaekel,
{\it Quantum vacuum fluctuations},
Compt. Rend. Acad. Sci. Ser. IV Phys. Astrophys. \textbf{2}, no.9, 1287-1298 (2001) [arXiv:quant-ph/0105053 [quant-ph]]


\bibitem{Sakharov:1967pk}
A.~D.~Sakharov,
{\it Vacuum quantum fluctuations in curved space and the theory of gravitation},
Dokl. Akad. Nauk Ser. Fiz. \textbf{177}, 70-71 (1967)

\bibitem{Streeruwitz:1975wzf}
E.~Streeruwitz,
{\it Vacuum fluctuations of a scalar field in an Einstein universe}, Phys. Lett. B \textbf{55}, 93-96 (1975)


\bibitem{Zeldovich:1971mw}
Y.~B.~Zeldovich and A.~A.~Starobinsky,
{\it Particle production and vacuum polarization in an anisotropic gravitational field},
Zh. Eksp. Teor. Fiz. \textbf{61}, 2161-2175 (1971)



\bibitem{Mainland:2018yqz}
G.~B.~Mainland and B.~Mulligan,
{\it How vacuum fluctuations determine the properties of the vacuum},
J. Phys. Conf. Ser. \textbf{1239}, no.1, 012016 (2019)


\bibitem{Kim:2016xvg}
S.~P.~Kim,
{\it Complex Effective Action and Schwinger Effect},
The Universe \textbf{4}, no.2, 8-16 (2016)
[arXiv:1611.08102 [hep-th]]

\bibitem{Srinivasan:1998fk}
K.~Srinivasan and T.~Padmanabhan,
{\it Facets of tunnelling: Particle production in external fields},
[arXiv:gr-qc/9807064 [gr-qc]]


\bibitem{Avramidi:1989ik}
I.~G.~Avramidi,
{\it Background field calculations in quantum field theory (vacuum polarization)},
Theor. Math. Phys. \textbf{79}, 494-502 (1989)





\bibitem{Schwinger1}
R. Brout, R. Parentani and Ph. Spindel, {\it Thermal properties of pairs produced by an electric field: A tunnelling approach}, Nuclear Physics B, \textbf{353}, no. 1, 209-236 (1991)


\bibitem{Sauter}
F.~Sauter, {\it Über das Verhalten eines Elektrons im homogenen elektrischen Feld nach der relativistischen Theorie Diracs}, Zeitschrift für Physik (1931)

\bibitem{Schwinger}
J.~S.~Schwinger, {\it On gauge invariance and vacuum polarization},  Phys. Rev. \textbf{82}, 664 (1951)


\bibitem{Parker:1968mv}
L.~Parker,
{\it Particle creation in expanding universes},
Phys. Rev. Lett. \textbf{21}, 562-564 (1968)

\bibitem{Parker:1969au}
L.~Parker,
{\it Quantized fields and particle creation in expanding universes. 1.},
Phys. Rev. \textbf{183}, 1057-1068 (1969)

\bibitem{QFTCS}
N.D. Birrell and P.C.W. Davies, {\it Quantum Fields In Curved Space}, Cambridge
University Press, (1982)
\bibitem{Hawking}
S. W. Hawking, {\it Particle Creation by Black Holes}, Commun. math. Phys. \textbf{43}, 199—220 (1975)

\bibitem{Hawking1}
S. W. Hawking, {\it Black Holes and Thermodynamics}, Phys. Rev. D \textbf{13} (1976) 191–197


\bibitem{Li:2016zyv}
Y.~Li, Y.~Dai and Y.~Shi,
{\it Pairwise mode entanglement in Schwinger production of particle-antiparticle pairs in an electric field}, Phys. Rev. D \textbf{95}, no.3, 036006 (2017)
[arXiv:1612.01716 [hep-th]]


\bibitem{Wu:2020dlg}
S.~M.~Wu and H.~S.~Zeng,
{\it Schwinger effect of Gaussian correlations in constant electric fields},
Class. Quant. Grav. \textbf{37}, no.11, 115003 (2020) [arXiv:2201.04001 [quant-ph]]

\bibitem{Ebadi:2014ufa} 
  Z.~Ebadi and B.~Mirza,
  {\it Entanglement Generation by Electric Field Background},
  Annals Phys.\  {\bf 351}, 363 (2014)
  [arXiv:1410.3130 [quant-ph]]

\bibitem{Adorno:2015ibo}
T.~C.~Adorno, S.~P.~Gavrilov and D.~M.~Gitman,
{\it Exactly solvable cases in QED with t-electric potential steps},
Int. J. Mod. Phys. A \textbf{32}, no.18, 1750105 (2017)
%doi:10.1142/S0217751X17501056
[arXiv:1512.01288 [hep-th]]

\bibitem{Fuentes:2010dt}
I.~Fuentes, R.~B.~Mann, E.~Martin-Martinez and S.~Moradi,
{\it Entanglement of Dirac fields in an expanding spacetime}
Phys. Rev. D \textbf{82}, 045030 (2010)
[arXiv:1007.1569 [quant-ph]]

\bibitem{Arias:2019pzy}
C.~Arias, F.~Diaz and P.~Sundell,
{\it De Sitter Space and Entanglement},
Class. Quant. Grav. \textbf{37}, no.1, 015009 (2020)
[arXiv:1901.04554 [hep-th]]


\bibitem{EE}
J. Maldacena and G. L. Pimentel, {\it Entanglement entropy in de Sitter space}, JHEP \textbf{1302}, 038 (2013) [arXiv:1210.7244 [hep-th]]
 
  
\bibitem{bell:2017}
S.~ Kanno and J.~Soda 
{\it Infinite violation of Bell inequalities in inflation
}, Phys.Rev.D \textbf{96}, 0211063  (2017)
[arXiv: 1705.06199 [hep-th]]





\bibitem{QC in deSitter}
J. Soda, S. Kanno and J. P. Shock,{\it Quantum Correlations in de Sitter Space}, Universe \textbf{3}, no. 1, 2 (2017)







\bibitem{Ebadi2015}
Z.~Ebadi and B.~Mirza, {\it Entanglement generation due to the
background electric field and curvature of space-time}, Int. J. Mod. Phys. A \textbf{30},1550031 (2015)


\bibitem{Grieninger:2023pyb}
S.~Grieninger, D.~E.~Kharzeev and I.~Zahed,
{\it Entanglement entropy in a time-dependent holographic Schwinger pair creation},
Phys. Rev. D \textbf{108}, no.12, 126014 (2023)
%doi:10.1103/PhysRevD.108.126014
[arXiv:2310.12042 [hep-th]].


\bibitem{Grieninger:2023ehb}
S.~Grieninger, D.~E.~Kharzeev and I.~Zahed,
{\it Entanglement in a holographic Schwinger pair with confinement},
Phys. Rev. D \textbf{108}, no.8, 086030 (2023)
doi:10.1103/PhysRevD.108.086030
[arXiv:2305.07121 [hep-th]].


\bibitem{HSSS}
S. Bhattacharya, S. Chakrabortty, H. Hoshino and S. Kaushal, \lq\lq {\it Background magnetic field and quantum correlations in the Schwinger effect} \rq\rq, Phys. Lett. B \textbf{811}, 135875 (2020) [arXiv:2005.12866 [hep-th]]


 \bibitem{SSSS}
M.~S.~Ali, S.~Bhattacharya, S.~Chakrabortty and S. Kaushal,
{\it``Fermionic Bell violation in the presence of background electromagnetic fields in the cosmological de Sitter spacetime''},
Phys. Rev. D \textbf{104}, no.12, 125012 (2021) 
[arXiv:2102.11745 [hep-th]]



\bibitem{SK}
S.~Kaushal,
``Schwinger effect and a uniformly accelerated observer,''
Eur. Phys. J. C \textbf{82}, no.10, 872 (2022) 
[arXiv:2201.03906 [hep-th]]







\bibitem{Plenio:2007zz} 
  M.~B.~Plenio and S.~Virmani,
  {\it An Introduction to entanglement measures},
  Quant.\ Inf.\ Comput.\  {\bf 7}, 1 (2007)
  [quant-ph/0504163]


\bibitem{Zyczkowski:1998yd} 
  K.~Zyczkowski, P.~Horodecki, A.~Sanpera and M.~Lewenstein,
{\it Volume of the set of separable states},
  Phys.\ Rev.\ A {\bf 58}, 883 (1998)
  [arXiv:quant-ph/9804024]

\bibitem{Martin}
M. B. Plenio, {\it Logarithmic Negativity: A Full Entanglement Monotone That is not Convex},  Phys. Rev. Lett. \textbf{95}, 119902 (2005) [	arXiv:quant-ph/0505071]

  \bibitem{Vidal:2002zz} 
  G.~Vidal and R.~F.~Werner,
  {\it Computable measure of entanglement},
  Phys.\ Rev.\ A {\bf 65}, 032314 (2002)
  [arXiv:quant-ph/0102117]
  
  \bibitem{wang}
X.~Wang and M.~M.~Wilde, {\it $\alpha$-logarithmic negativity
}, Phys.Rev.A 102 (2020) 3, 032416 [arXiv:  1904.10437 [quant-ph]]
  

\bibitem{Plenio:2005} 
  M.~B.~Plenio,
  {\it Logarithmic negativity: a full entanglement monotone that is not convex},
  Phys.\ Rev.\ Lett.\  {\bf 95}, 090503 (2005)
  
  \bibitem{Calabrese:2012nk} 
  P.~Calabrese, J.~Cardy and E.~Tonni,
  {\it Entanglement negativity in extended systems: A field theoretical approach}
  J.\ Stat.\ Mech.\  {\bf 1302}, P02008 (2013)
  [arXiv:1210.5359 [cond-mat.stat-mech]]

\bibitem{Nishioka:2018khk} 
  T.~Nishioka,
  {\it Entanglement entropy: holography and renormalization group},
  Rev.\ Mod.\ Phys.\  {\bf 90}, no. 3, 035007 (2018)
  [arXiv:1801.10352 [hep-th]]

\bibitem{jmath}
D. Bruß, {\it Characterizing entanglement}, Journal of Mathematical Physics \textbf{43}, 4237 (2002) [arXiv: quant-ph/0110078].

 \bibitem{NielsenChuang} 
	M.~A.~Nielsen and I.~L.~Chuang (2010), {\it Quantum Computation and Information Theory} (Cambridge university press, UK)

 \bibitem{seprability}
M.~Horodecki, P.~Horodecki and R.~Horodecki, {\it Separability of mixed states: Necessary and suffcient conditions}, Phys. Lett.A \textbf{223}, (1996) [arXiv:quant-ph/9605038]


\bibitem{bellmix}
S. Popescu and D. Rohrlich,
{\it Generic quantum nonlocality},
Physics Letters A,
\textbf{166}, no. 5,
293-297, (1992)


\bibitem{monogamy1}
V.~Coffman, J.~Kundu and W.~K.~Wootters,
{\it Distributed entanglement},
Phys. Rev. A \textbf{61}, 052306 (2000)
[arXiv:quant-ph/9907047 [quant-ph]]

\bibitem{monogamy2}
T.~J.~Osborne and F.~Verstraete,
{\it General Monogamy Inequality for Bipartite Qubit Entanglement},
Phys. Rev. Lett. \textbf{96}, no.22, 220503 (2006)



\bibitem{monogamy3}
E.~Martin-Martinez and I.~Fuentes,
{\it Redistribution of particle and anti-particle entanglement in non-inertial frames},
Phys. Rev. A \textbf{83}, 052306 (2011)
%doi:10.1103/PhysRevA.83.052306
[arXiv:1102.4759 [quant-ph]]

\bibitem{AS}
	M.~Abramowitz and I.~Stegun,
{\it Handbook of Mathematical Functions with Formulas, Graphs, and Mathematical Tables}, National Bureau of Standards (USA) (1964)



  



\end{thebibliography}
\end{document}